\documentclass[longauth]{aa}  
\usepackage{graphicx}
\usepackage{txfonts}
\usepackage{color}
\usepackage{xcolor}
\definecolor{byzantine}{rgb}{0.74, 0.2, 0.64}

\begin{document} 

   \title{The PLATO field selection process}

   \subtitle{I. Identification and content of the long-pointing fields}

   \author{
   V.~Nascimbeni\thanks{E-mail: valerio.nascimbeni@inaf.it} \inst{1,2} \and
   G.~Piotto\inst{2,1} \and
   A.~B\"orner\inst{3} \and
   M.~Montalto\inst{2,1} \and
   P.~M.~Marrese\inst{4,5} \and
   J.~Cabrera\inst{6} \and
   S.~Marinoni\inst{4,5} \and
   C.~Aerts\inst{7,8,9} \and
   G.~Altavilla\inst{4,5} \and
   S.~Benatti\inst{10} \and
   R.~Claudi\inst{1} \and
   M.~Deleuil\inst{11} \and
   S.~Desidera\inst{1} \and
   M.~Fabrizio\inst{4,5} \and
   L.~Gizon\inst{12,13,14} \and
   M.~J.~Goupil\inst{15} \and
   V.~Granata\inst{2,1} \and
   A.~M.~Heras\inst{16} \and
   D.~Magrin\inst{1} \and
   L.~Malavolta\inst{2,1} \and
   J.~M.~Mas-Hesse\inst{17} \and
   S.~Ortolani\inst{2,1} \and
   I.~Pagano\inst{18} \and
   D.~Pollacco\inst{19,20} \and
   L.~Prisinzano\inst{11} \and
   R.~Ragazzoni\inst{2,1} \and
   G.~Ramsay\inst{21} \and
   H.~Rauer\inst{6,22,23} \and
   S.~Udry\inst{24}
          }

   \institute{
   INAF -- Osservatorio Astronomico di Padova, vicolo dell'Osservatorio 5, 35122 Padova, Italy \and
   Dipartimento di Fisica e Astronomia ``Galileo Galilei'', Universit\`a degli Studi di Padova, Vicolo dell'Osservatorio 3, 35122 Padova, Italy \and
   Deutsches  Zentrum  f\"ur  Luft-  und  Raumfahrt  (DLR),  Institut  f\"ur  Optische  Sensorsysteme,  Rutherfordstra{\ss}e  2, 12489 Berlin-Adlershof, Germany \and
   INAF -- Osservatorio Astronomico di Roma, Via Frascati, 33, 00078 Monte Porzio Catone (RM), Italy \and
   SSDC-ASI, Via del Politecnico, snc, 00133 Roma, Italy \and
   Deutsches Zentrum f\"ur Luft- und Raumfahrt (DLR), Institut f\"ur Planetenforschung, Rutherfordstra{\ss}e 2, 12489 Berlin-Adlershof, Germany\and
   Institute of Astronomy, KU Leuven, Celestijnenlaan 200D, 3001, Leuven, Belgium \and
   Department of Astrophysics, IMAPP, Radboud University Nijmegen, 6500 GL, Nijmegen, The Netherlands  \and
   Max Planck Institute for Astronomy, Koenigstuhl 17, 69117, Heidelberg, Germany \and
   INAF -- Osservatorio Astronomico di Palermo, Piazza del Parlamento 1, 90134 Palermo, Italy \and
   Aix-Marseille Universit\'e, CNRS, CNES, Laboratoire d’Astrophysique de Marseille,Technop\^{o}le de Marseille-Etoile, 38, rue Fr\'ed\'eric Joliot-Curie, 13388 Marseille cedex 13, France \and
   Max-Planck-Institut f\"ur Sonnensystemforschung, Justus-von-Liebig-Weg~3, 37077~G\"ottingen, Germany \and
   Institut f\"ur Astrophysik, Georg-August-Universit\"at G\"ottingen, Friedrich-Hund-Platz~1, 37077~G\"ottingen, Germany \and
   Center for Space Science, NYUAD Institute, New York University Abu Dhabi, Abu Dhabi, UAE \and
   LESIA, CNRS UMR 8109, Universit\'e Pierre et Marie Curie, Universit\'e Denis Diderot, Observatoire de Paris, 92195 Meudon, France \and
   European Space Agency (ESA), European Space Research and Technology Centre (ESTEC), Keplerlaan 1, 2201 AZ Noordwijk, The Netherlands \and
   Centro de Astrobiolog\'{\i}a (CSIC--INTA), Depto. de Astrof\'{\i}sica, 28692 Villanueva de la Ca\~nada, Madrid, Spain\and
   INAF -- Osservatorio Astrofisico di Catania, Via S. Sofia 78, 95123, Catania, Italy \and
   Department of Physics, University of Warwick, Gibbet Hill Road, Coventry CV4 7AL, UK \and
   Centre for Exoplanets and Habitability, University of Warwick, Gibbet Hill Road, Coventry CV4 7AL, UK\and
   Armagh Observatory \& Planetarium, College Hill, Armagh, BT61 9DG, UK \and
   Zentrum f\"ur Astronomie und Astrophysik, TU Berlin, Hardenbergstra{\ss}e 36, 10623 Berlin, Germany \and
   Institute of Geological Sciences, Freie Universit\"at Berlin, Malteserstra{\ss}e 74-100, 12249 Berlin, Germany \and
   Observatoire de Gen\`eve, Universit\'e de Gen\`eve, Chemin Pegasi 51, 1290 Sauverny, Switzerland 
   }

   \date{Received 19 September 2021 / Accepted 26 October 2021}

  \abstract{PLAnetary Transits and Oscillations of stars (PLATO) is an ESA M-class satellite planned for launch by the end of 2026 and dedicated to the wide-field search of transiting planets around bright and nearby stars, with a strong focus on discovering habitable rocky planets hosted by solar-like stars. The choice of the fields to be pointed at is a crucial task since it has a direct impact on the scientific return of the mission. In this paper, we describe and discuss the formal requirements and the key scientific prioritization criteria that have to be taken into account in the Long-duration Observation Phase (LOP) field selection, and apply a quantitative metric to guide us in this complex optimization process. We identify two provisional LOP fields, one for each hemisphere (LOPS1 and LOPN1), and we discuss their properties and stellar content. While additional fine-tuning shall be applied to LOP selection before the definitive choice, which is set to be made two years before launch, we expect that their position will not move by more than a few degrees with respect to what is proposed in this paper.}


   \maketitle
%


\section{Introduction}

Since the dawn of modern exoplanetary science in the 1990s, one of its most ambitious aspirations has always been the discovery and characterization of ``true'' Earth analogs, that is rocky planets orbiting within the habitable zone (HZ) of solar twins. By looking at the distribution of the currently known exoplanets, it is evident that only a handful of candidates approach that sweet spot in the parameter space, and that all of them actually miss a reliable mass estimate, being hosted by stars too faint to be properly investigated through ultra-high-precision radial velocities (RVs). After the pioneering results from the CoRoT satellite (built by an international consortium led by CNES; \citealt{Auvergne2009}) and from the very successful NASA missions Kepler \citep{Borucki2010} and TESS \citep{Ricker2015} demonstrated that space-based, wide-field photometry is extremely effective in detecting transiting planets, and new-generation ultra-stable spectrographs based on the HARPS/HARPS-N legacy \citep{Mayor2003,Cosentino2012} such as ESPRESSO \citep{Pepe2021} and the planned HIRES@E-ELT \citep{Marconi2021} are getting closer and closer to the 10~cm/s RV precision level  needed to confirm true Earths. In this context, the ESA M-class mission PLAnetary Transits and Oscillation of Stars (PLATO; \citealt{Rauer2014}), planned for launch in 2026, is designed to push the current technology at its extremes in combined terms of photometric precision and the overall field of view (FOV). The latter is clearly linked to the number of available bright, solar-type main-sequence stars, which in turn will allow not only a proper follow-up by ground-based facilities, but also the extraction of stellar parameters (including age) with an unprecedented accuracy from the asteroseismological analysis of the light curves themselves.

\begin{figure}
    \centering
    \includegraphics[width=0.95\columnwidth]{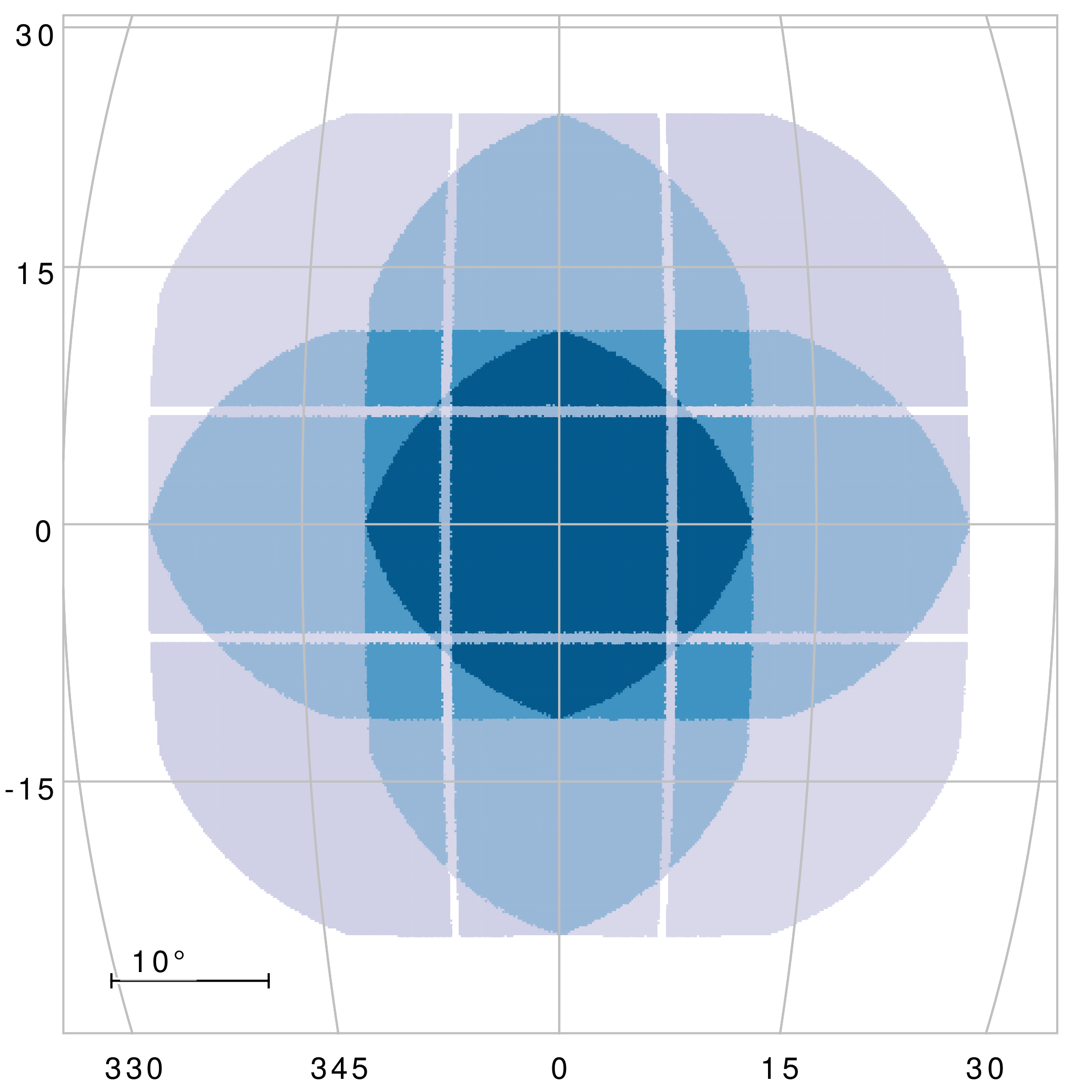}
    \caption{Geometry of the PLATO field FOV, here it is centered at the origin (0,0) of a generic spherical reference frame (units are degrees; projection is orthographic). The number of ``normal'' cameras covering a given line of sight is color-coded. The four blue shades, from dark to light, map regions observed with 24, 18, 12, six cameras (corresponding to four, three, two, and one group(s) of six co-pointing telescopes each, respectively).}
    \label{fov}
\end{figure}

The optical design of PLATO is based on 24 identical 20cm-class telescopes with an entrance pupil of 12-cm (dubbed ``normal'' cameras) optimized for a single 450-1000~nm band pass, each having a net circular FOV of about $1037$~$\textrm{deg}^2$ \citep{Ragazzoni2016}. Those cameras are not all co-pointing, but rather arranged in four groups of six cameras each, with the line of sight of each group offset by $9.2^\circ$ with respect to the satellite bore-sight, along perpendicular directions (Fig.~\ref{fov}). Such a pattern is nearly invariant to $90^\circ$ rotations around the bore-sight axis, which are mandatory at a three-month cadence to keep the solar panels of PLATO optimally oriented toward the Sun. The final result, after only accounting for the parts of the focal planes covered by CCDs, is an overall FOV of about 2132~$\textrm{deg}^2$, covered by a variable amount of 24, 18, 12, and six telescopes\footnote{The nominal covered FOV depends upon the geometry of the focal plane and the optical quality of the centered optical system; it can be found in detail elsewhere \citep{Pertenais2021,Magrin2021} and it is expected to slightly vary in actual cameras in a manner that is expected not to affect the discussion hereafter.}. As a consequence of this overlapping design, stars within the inner region of the PLATO field (PF) are monitored with a much higher photometric precision than the peripheral ones. In addition to the normal cameras, two more similar telescopes (dubbed "fast'' cameras) will monitor the inner region by adding color information and a much faster sampling cadence (2.5~s versus~25~s).

Such a huge field, with a roundish square footprint of about $49^\circ\times 49^\circ$ overall, will simultaneously image about 5\% of the whole sky. The observing strategy of PLATO, which has yet to be finalized, will probably be two-staged. The current operation baseline scenario includes two pointings that are to be monitored during the Long-duration Observation Phase (LOP fields) of two years each. The minimum duration of a LOP field is one year, and LOP fields can be complemented by shorter pointings during the ``Step and stare'' Observation Phase (SOP fields), which can expand the total areas surveyed to 40\% of the sky.
It is worth emphasizing that the observing strategy has not been formally fixed yet, nor the number or the exact duration of the SOP and LOP fields. Nonetheless, the LOP fields are recognized as being the most important ones to achieve the scientific objectives of the mission, and will be very likely the first one(s) to be observed. Hence, the choice of the LOP fields is both of crucial importance and also the most urgent one, and we focus the present work on it. This urgency has been expressed not only for the PLATO consortium, but also for the exoplanetary and stellar communities at large since the final choice will have a large impact on many ongoing research programs and on the planning of target characterization tools and follow-up campaigns. For example, these include the monitoring of the level of stellar activity of the high-priority targets.

In this paper we present all of the formal requirements, scientific criteria, and prioritization algorithms, leading to a preliminary (yet close to final) choice for the two LOP fields, one in the northern ecliptic hemisphere (LOPN1: LOP field North~1) and one in the southern one (LOPS1: LOP field South~1). This paper is also intended to stimulate a discussion within the community to converge on a final choice for the LOP fields. A second, future paper will be devoted to the fine-tuning of the LOP fields and to the selection of candidate SOP fields. The present paper is organized as follows. In Section 2 we describe the geometry and the general constraints on the field selection problem. In Section 3 we define a quantitative metric to prioritize a given LOP field according to its scientific value, and apply this metric to a grid of possible pointings in Section 4. Finally, we identify and characterize the content of LOPN1 and LOPS1 in Section 5 and 6, respectively. Some final remarks and future perspectives are summarized in Section 7. For the sake of clarity, a glossary with the most used acronyms in this work is shown in Table~\ref{table:glossary}.

\begin{table}\centering
\caption{Glossary of acronyms used throughout this article.}
\begin{tabular}{lp{5cm}}
\hline
Acronym & Description \\
\hline\hline
asPIC & All-sky PLATO Input Catalog \citep{Montalto2021}\\
CVZ & Continuous Viewing Zone \\
DEB & Detached Eclipsing Binary\\
DIA & Difference Image Analysis \\
EB & Eclipsing Binary \\
FOV & Field Of View \\
GC & Globular Cluster \\
GO & PLATO Guest Observing program\\
HZ & Habitable Zone \\
LMC & Large Magellanic Cloud\\
LOP & Long-duration Observation Phase \\
LOPN & LOP field North\\
LOPN0 & Obsolete LOPN proposal \citep{Nascimbeni2016} \\
LOPN1 & Current LOPN proposal (this work) \\
LOPS & LOP field South\\
LOPS0 & Obsolete LOPS proposal \citep{Nascimbeni2016} \\
LOPS1 & Current LOPS proposal (this work) \\
NPF & Obsolete acronym for LOPN \\
NSR & Noise-to-Signal ratio \\
OC & Open Cluster \\
P/L & PLATO Payload \\
PF & PLATO field \\
PIC & PLATO Input Catalog \citep{Montalto2021} \\
PLATO & PLAnetary Transits and Oscillations of stars \citep{Rauer2014} \\PPT & PLATO Performance Team \\
PSF & Point Spread Function \\
RV & Radial velocity \\
S/N & Signal-to-Noise ratio \\
SOP & Step and stare Observation Phase \\
SPF & Obsolete acronym for LOPS \\
SRD & PLATO Scientific Requirements Document \\
SRJD & PLATO Scientific Requirements Justification Document \\
\hline
\end{tabular}\label{table:glossary}
\end{table}

\section{The field selection problem}

The present work is the first attempt to identify the LOP fields using a consistent and quantitative approach. The coordinates of the SOP and LOP fields previously selected, as reported in some papers (such as \citealt{Nascimbeni2016}; \citealt{Miglio2017}; \citealt{Marchiori2019}; \citealt{Montalto2021}), all centered at Galactic latitude $|b|=30^\circ$, were merely a working hypothesis to verify their compliance with the PLATO Science Requirements Document (SRD; v8.0). The definitive choice for the first observing field will be formally delivered two years before launch with the possibility of fine tuning the target list up to nine months before launch. In this section, we review the starting points of this selection process. 

Due to telemetry constraints, PLATO will not be able to download time series of full-frame images. Rather, it will download light curves  of preselected targets, and ``imagettes'' for a fraction of them, drawn from the PLATO Input Catalog (PIC; \citealt{Montalto2021}). 
This implies, as it will be further discussed in Section~\ref{metric}, that any field-level prioritization scheme must be based on a target-level scheme. With only a few exceptions, targets belong to four formally defined (SRD) samples\footnote{For historical reasons, the sample P3 has been eliminated, but the numbering of the PLATO samples was left unchanged.}:
\begin{itemize}
    \item P1: main-sequence and subgiants stars in the F5-K7 range of spectral type, brighter than $V=11$, which are to be observed by PLATO with a photometric noise\footnote{The noise limit in this context is established for the random noise component alone. As required by the SRD, the systematic noise component is, at most, one-third of the random noise at $V=11$.}  smaller than 50~ppm in one hour;
    \item P2: same as P1, but brighter than $V=8.5$; 
    \item P4: cool late-type dwarfs (M dwarfs) later than K7 and brighter than $V<16$;
    \item P5: main-sequence and subgiants stars from F5 to late K spectral types, brighter than $V=13$ and with no constraints on photometric noise.
\end{itemize}
While the PIC will contain only the targets that are to be observed by PLATO on the final LOP and SOP fields, an all-sky version of it, including all the stars compliant with the P1-P2-P4-P5 definition (except the requirements on photometric noise), is also available (asPIC; \citealt{Montalto2021}) and will be the starting point of our simulations. Additional catalogs  are being compiled in parallel for operational purposes (including instrumental calibration, validation, and fine pointing).

The PLATO scientific requirements for the PIC are described in the SRD for the samples mentioned above; we refer the reader to \citet{Montalto2021} for a detailed overview\footnote{A new paper by Rauer et al., giving a general status update of the PLATO mission including the target and field selection work, is currently in preparation.}.
Throughout this work, we mostly focus on prioritizing the P1 sample because it is recognized as the ``backbone of the PLATO mission and must be considered as the highest priority objective'' (SRD), and the field choice should consequently be driven by it.

The field selection process involves a complex optimization task to merge several (often competing) constraints and prioritization criteria of both an engineering and scientific nature. We discuss those criteria by splitting them into four main classes as detailed in the following subsections.

\begin{figure*}
    \centering
    \includegraphics[width=1.8\columnwidth]{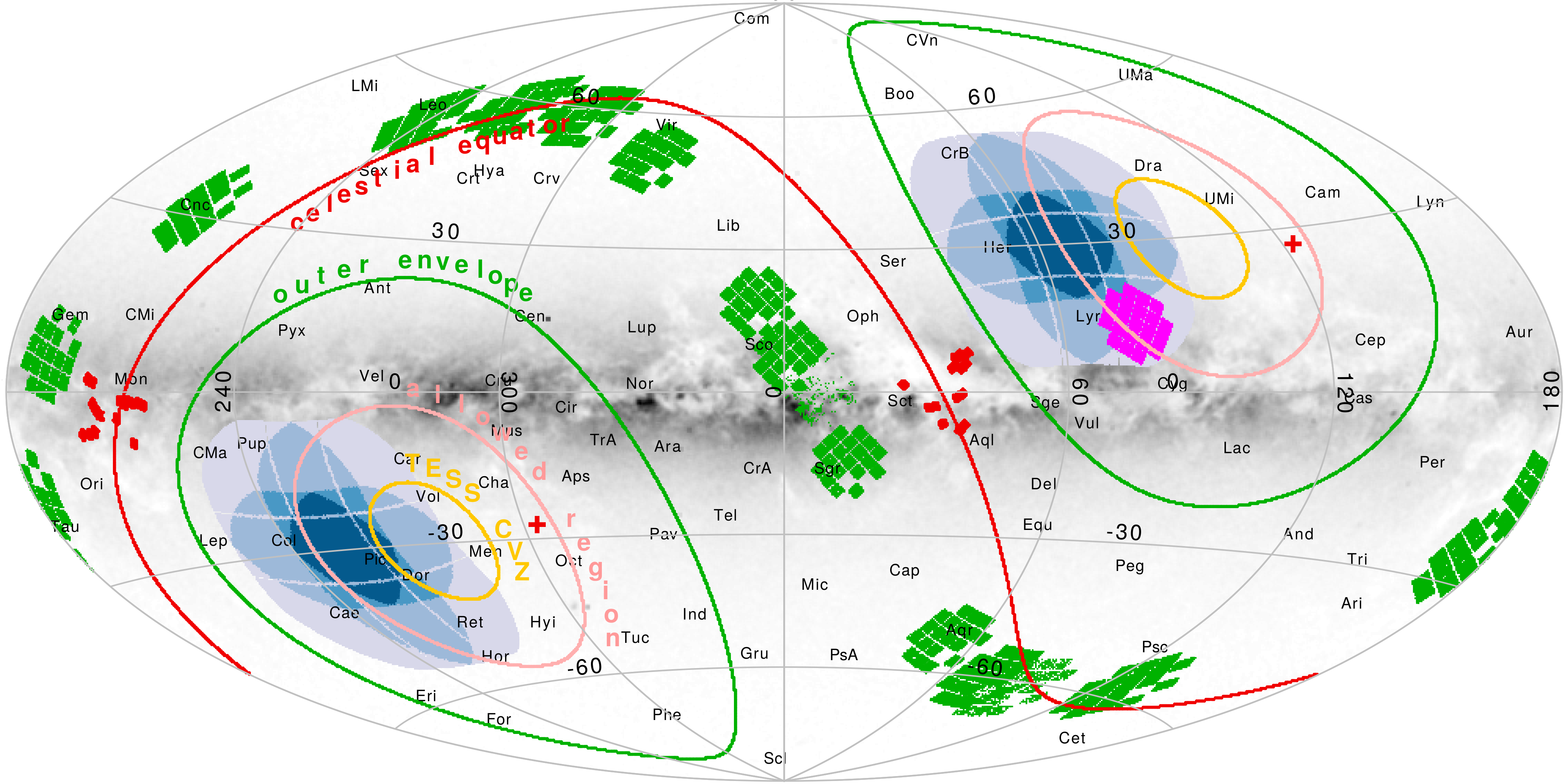}
    \caption{All-sky Aitoff projection in Galactic coordinates, showing the formal constraints for the selection of the PLATO LOP fields and the synergies with other transit-finding missions. 
    The two pink circles represent the $|\beta|>63^\circ$ technical requirement for the center of the LOP fields (``allowed region''), implying that the overall envelopes of every allowed field choice are two ecliptic caps at $|\beta|\gtrsim38^\circ$ (green circles). As a reference, the previous provisional choices for the LOP fields, here called LOPN0 and LOPS0, are plotted with blue shades according to the number of co-pointing cameras (refer to Fig.~\ref{fov} for details). The footprints of CoRoT (red), Kepler (magenta), and K2 (green) are over-plotted together with the TESS continuous viewing zone at $|\beta|\gtrsim78^\circ$ (yellow circle). The background gray layer is color-coded according to the areal density of $G<13.5$ stars from Gaia EDR3. The celestial equator and poles are marked with a red circle and crosses, respectively.}
    \label{full_sky}
\end{figure*}

\subsection{Formal requirements}\label{requirements}

By ``formal requirements'', we mean mandatory requirements that are listed and justified in the PLATO SRD and Scientific Requirements Justification Document (SRJD), respectively. The most relevant ones for the field selection task are the following: 

\begin{enumerate}
    \item \emph{Geometrical constraints.} The center of the LOP fields shall lie within the two caps at ecliptic latitude $|\beta| > 63^\circ$ (SRD) to allow for a proper orientation of the spacecraft throughout the year. We refer to these caps as ``allowed regions", and to the LOP fields within them as North and South PLATO LOP field (LOPN, LOPS), respectively. As this requirement holds just for the center of the fields, part of them can extend outside of the allowed region anyway (see below); 
    \item \emph{Target counts.} The number of observed targets included in the PLATO stellar samples as defined by the SRD must pass a specified threshold for each of the four stellar samples monitored by PLATO (P1, P2, P4, and P5). For the Stellar Sample 1 (P1), the LOP fields combined must contain at least $15\,000$ (goal: $20\,000$) F5-K7 dwarfs and subgiants brighter than $V=11$ and be observed with a photometric precision better than 50~ppm in one hour. Other analogous thresholds are set for samples P2, P4, and P5, but they can be safely neglected in this context because they are always met for every possible choice for the LOP fields (see Section~\ref{content}).
\end{enumerate}    
    
The $|\beta| > 63^\circ$ constraint is visualized with an Aitoff projection in Fig.~\ref{full_sky} (pink lines); the two allowed regions combined represent about $1-\sin (63^\circ)\simeq 11\%$ of the whole sky area. It is worth emphasizing that this requirement holds just for the centers of the PF, that is, the outer envelope of every possible PF choice can reach as far as the $|\beta| \simeq 38^\circ$ (green lines in Fig.~\ref{full_sky}). In other words, in principle, approximately $1-\sin (38^\circ)\simeq 38\%$ of the whole sky could be accessible accounting for this constraint. 
The two allowed regions are located at high absolute declinations ($|\delta| > 40^\circ$), with the celestial poles being located within them. On the other hand, they span a wide range of Galactic latitudes, from being nearly tangent to the Galactic plane at $b\sim 2.8^\circ$ up to $b\sim 56.8^\circ$. Once the full envelope is taken into account, that implies that every possible Galactic environment could be potentially probed by a LOP field. 





\subsection{Prioritization criteria} \label{criteria}

A second class of criteria are based on scientific and practical considerations. While not formally mandatory, these criteria can be applied to prioritize those fields which, while fulfilling the formal requirements, also maximize the scientific output of the mission and minimize the follow-up effort. Among those, we would like to highlight the following:

\begin{enumerate}    
    \item \emph{Stellar contamination.} Moving the PF closer to the Galactic plane boosts the stellar counts for some PLATO stellar samples (including P1; see Section~\ref{grid}) and also greatly increases the density and combined flux of background contaminants, which can be a source of both additional photometric noise ({photometric contamination}) and of an increased fraction of false positives mostly given by blended and grazing eclipsing binaries ({astrophysical contamination}; \citealt{Almenara2009}; \citealt{Bryson2013}; \citealt{Fressin2013}) combined with more exotic scenarios (such as the blending of genuine planetary transits or eclipsing binaries for which only the secondary eclipses are visible; \citealt{Santerne2013}). PLATO, similarly to TESS, is particularly sensitive to both, owing to its large pixel scale ($\sim 15''$; \citealt{Rauer2014}).
    Contamination has, of course, an important impact on the full validation and confirmation chain, both based on the PLATO photometry itself and on the follow-up observations.
    \item \emph{Follow-up resources.} The whole follow-up process is an important part of the mission, in particular to confirm the planetary nature of the candidates and to measure their masses through high-precision radial velocities (RVs). The location of the LOP fields has some obvious and important implications for the available ground-based facilities (based on the celestial coordinates of the targets and their annual visibility) that will be assessed in Section~\ref{fu}. More subtly, the fraction of false positives and the actual content of the target list in terms of astrophysical parameters also have a profound impact on the follow-up (see Sec.~\ref{content});  
        \item \emph{Special targets of interest.} Especially at the fine-tuning stage, it makes sense to check whether specific objects of high scientific interest land on the actual CCD collecting area or if this could be made feasible by only minor adjustments to the pointing. This includes, for instance, nearby open clusters, known transiting planets, and very bright solar analogs. Some of these objects will be discussed in Section~\ref{content}.
    \item \emph{Synergy with other missions.} Other space missions, most notably Kepler and TESS, have already monitored a significant fraction of the sky in search of planetary transits. While the main goal of PLATO is to exploit its unprecedented level of photometric precision for the detection and characterization of true Earth analogs, it is worth investigating whether interesting additional science may come up by the overlap with other observed fields.
\end{enumerate}

As for the synergies with other exoplanetary missions, we note that all the K2 and CoRoT \citep{Auvergne2009} fields lie very close to the ecliptic and celestial equator, respectively, that is completely out of reach of the LOP envelope (Fig.~\ref{full_sky}). On the other hand, the center of the Kepler Field is located within the LOPN allowed region, and the TESS continuous viewing zone (CVZ) at $|\beta|\gtrsim78^\circ$ is of course centered on the ecliptic poles, that is, it at least partly overlaps with every technically acceptable LOP field choice. The James Webb Space Telescope (JWST) ``strict'' CVZ at $|\beta|>85^\circ$ \citep{Gardner2006} is a much smaller cap. Nevertheless, JWST will be able to observe every source at $|\beta|>45^\circ$ for at least 200~days per year in one continuous chunk, therefore essentially the vast majority of targets within the LOP outer envelope can be optimally monitored by JWST.

\begin{figure*}\label{fig:fu}
    \centering
    \includegraphics[width=0.95\columnwidth]{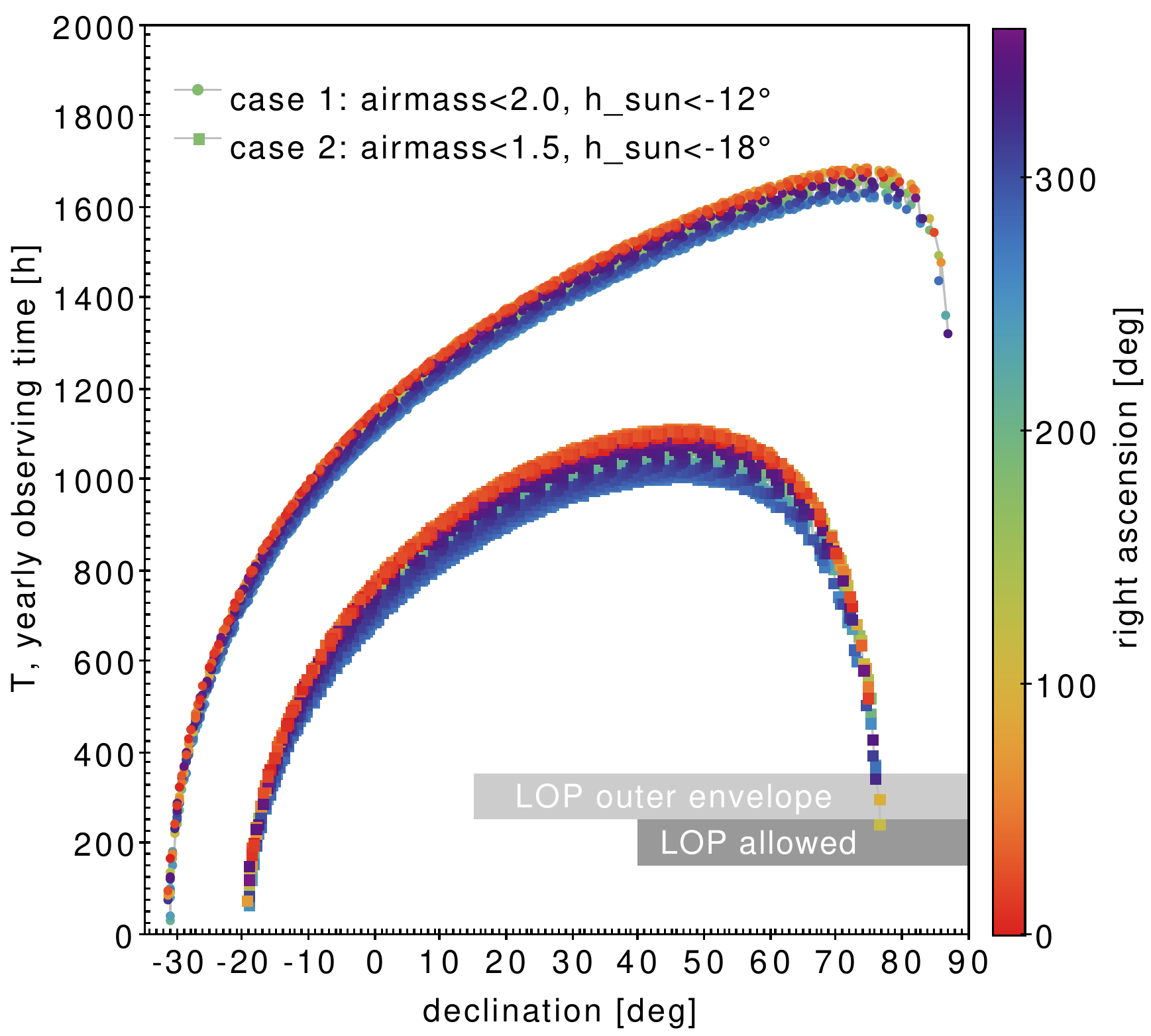}\hspace{10mm}
    \includegraphics[width=0.95\columnwidth]{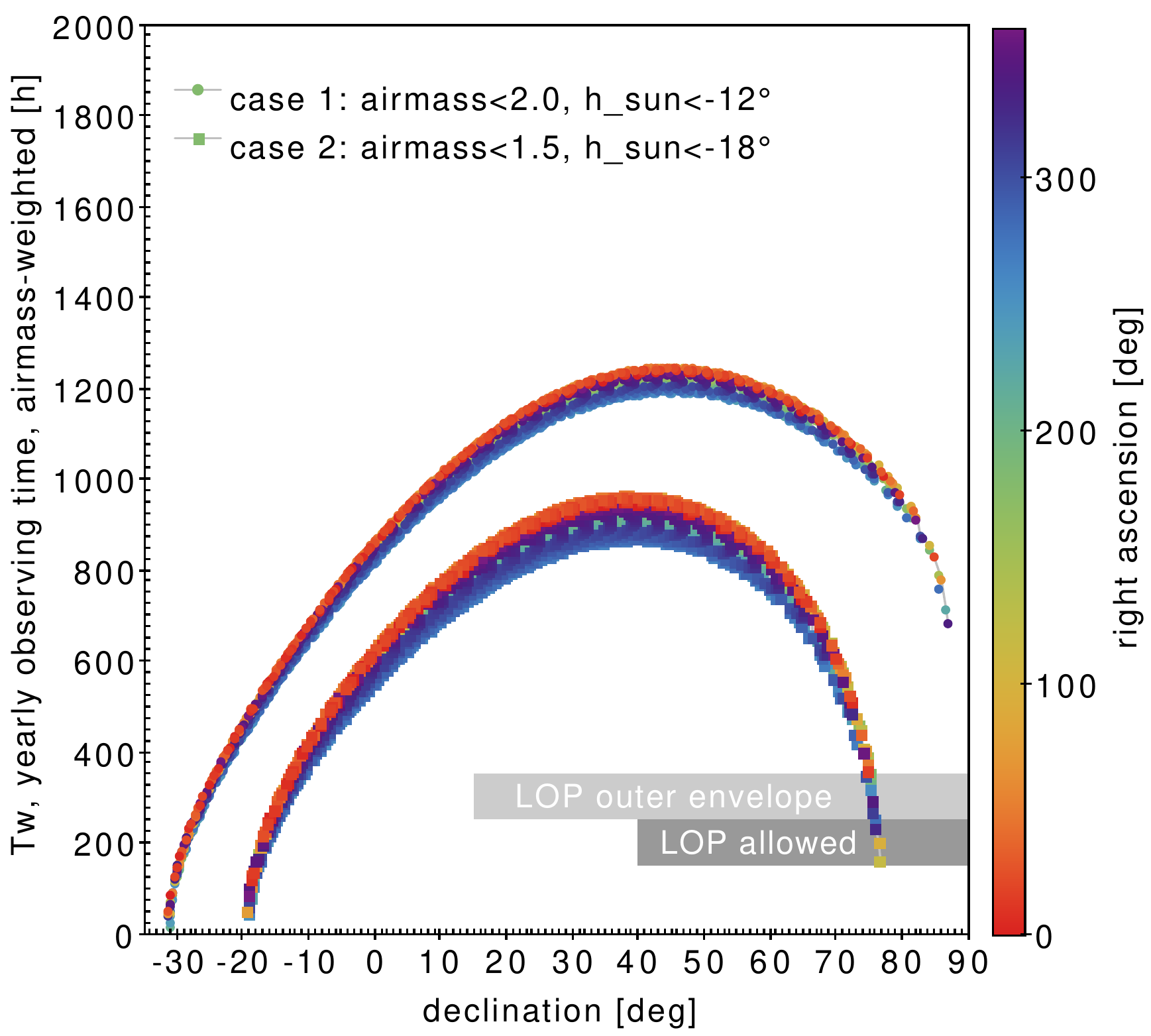}
    \caption{\emph{Left panel:} Integrated yearly observing time $T$ in hours of a given star from a hypothetical astronomical observatory at latitude $\varphi=28.5^\circ$, as a function of its equatorial coordinates $\alpha$ (color scale) and $\delta$ (horizontal axis). The two curves refer to two different sets of thresholds imposed on the maximum airmass $X$ and the maximum solar elevation $h_\odot$, as in legend. The declination range spanned by the $|\beta|>38^\circ$ ``outer envelope'' and $|\beta|>63^\circ$ ``allowed region'' (see Fig.~\ref{full_sky}) for the LOPN field is highlighted with gray boxes. Choosing an observatory with the same $|\varphi|$ but in the southern hemisphere would yield the same results but with the $\delta$ sign reversed. \emph{Right panel:} Same as above, but now the integrated time $T_w$ is weighted by $1/X$, as a more effective follow-up metric.}
    \label{yearly}
\end{figure*}

\subsection{Ease of follow-up}\label{fu}

The impact of the field choice on the follow-up effort, from a purely observational point of view, can be quantitatively addressed by computing the total amount of observing time $T$ available for a given target and from a given ground-based site as a function of its coordinates, under reasonable assumptions. Most top-class facilities that are or will be available for the PLATO follow-up are located at tropical or subtropical latitudes (for instance, La Palma at $\varphi\sim28^\circ.5$, Paranal at $\varphi\sim-24^\circ.5$, La Silla at $\varphi\sim-29^\circ$, and Mauna Kea at $\varphi\sim20^\circ$). The ``net'' observing time must of course be computed with reasonable constraints based on the airmass $X$ and on the apparent elevation of the Sun $h_\odot$, with 
\begin{enumerate}
    \item most observations being carried out at $X<2$, $h_\odot<-12^\circ$ (hereafter {case 1}), and
    \item very challenging and/or faint targets requiring much safer limits at $X\lesssim1.5$, $h_\odot<-18^\circ$ (hereafter {case 2}).
\end{enumerate}
Between these two extremes lies the intermediate case of ultra-high-precision RVs, usually requiring $X\lesssim1.5$, $h_\odot<-12^\circ$ to maximize the signal-to-noise ratio (S/N) and mitigate telluric contamination \citep{Cunha2014}.

As an exercise, we took a hypothetical astronomical observatory at $\varphi = 28^\circ.5$ as a reference site\footnote{Of course, choosing a southern observatory with a similar $|\varphi|$ and reversing the declination sign would yield the same results for the LOPS choice, so we refer to $|\delta|$ hereafter.}, computed a uniform HEALPix level-4 grid in equatorial coordinates $(\alpha, \delta)$ (i.e., $3\,072$ grid points; \citealt{Gorski2005}), and integrated over a full solar year the total number of hours at which a star at given $(\alpha, \delta)$ can be observed below the thresholds on $X$ and $h_\odot$ mentioned above. The simulation was performed through the SLALIB library \citep{Wallace2005} and its output is plotted in the left panel of Fig.~\ref{yearly} as a function of $\delta$ (horizontal axis) and $\alpha$ (color scale). As expected, declination is the key variable, with the yearly observing time reaching the maximum at $|\delta|\simeq 75^\circ$ for case 1 (upper sequence) and  $|\delta|\simeq 45^\circ$ for case 2 (lower sequence). The dependence of $T$ on right ascension $\alpha$ is quite weak ($\lesssim 15\%$, everything else being equal) because with such constraints, the nightly observing window is mostly limited by the net arc length of the sky track (set by $X$) rather than by the twilight-to-twilight night duration (set by $h_\odot$). As expected, $T$ drops when approaching the celestial pole, which is always at $X>2$ at a $\varphi = 28^\circ.5$ observing site; this drop crosses 50\% of the $T$ peak at $|\delta|\simeq 87^\circ$ and $|\delta|\simeq 75^\circ$ for case 1 and 2, respectively. In other words, the LOP fields covering those high-declination caps should be given lower priority in the field selection process because they severely limit the follow-up observations.   

Taking a step further, as a more effective metric, we could compute the same quantity by giving higher priority to observations carried out at lower airmass. This can be achieved by defining the airmass-weighted observing time $T_w$ as $T$ weighted by a $1/X$ factor (Fig.~\ref{yearly}, right panel). By looking at the general trend, there is clearly a ``sweet spot'' with more then 800~h/year available at $20^\circ\lesssim|\delta|\lesssim 60^\circ$, where the follow-up can be carried out at a much higher efficiency. Coincidentally, this is approximately the declination range spanned by our initial LOP field choices LOPS0 and LOPN0\footnote{In some previous works, the LOPS0 and LOPN0 field were referred to as South and North PLATO Fields (SPF and NPF), respectively.}  (Fig.~\ref{full_sky}). By setting a granularity of 10~minutes as minimum on-target time per night for case 2, within the same declination range, we get an approximate number of $\sim 250$ available nights per year, enabling an optimal phase sampling of the RV signal from HZ planets.
Future simulations will be focused on the actual observing sites involved in the PLATO follow-up program and will include additional variables not considered here, such as Moon constraints and weather statistics.

\subsection{Field rotation}

The PLATO attitude will be defined not only by a pointing direction, but also by a rotation angle around the $Z$ axis of the payload (P/L) module. Given a pointing direction, different rotation angles will result in a different set of stars observed. Some of the field selection criteria are only weakly dependent on the choice of the rotation angle (e.g., the P1 counts), whereas some are strongly dependent (e.g., visibility of individual targets close to the edges of the field of view). The choice of the rotation angle determines the angle between the solar panels of the spacecraft and the Sun. Therefore, there is a dependency between the choice of the rotation angle at the start of the observations and the time at which it is necessary to make the first 90 degree roll to keep the action of the sun shield within the requirements.
The constraints related to the spacecraft such as: power produced by the solar panels at a given sun incidence angle, allowed ranges for early or late 90 degree rolls along the year compliant with the gap requirements, stray light requirements, etc., are not consolidated at this stage of the project and therefore we cannot define reliable criteria for the choice of the optimal rotation angle now. The mission team is working toward a better understanding of the design in order to provide enough information to the PLATO Science Working Team to make a choice.

\section{The prioritization metric}\label{metric}

As a starting point for the identification of the LOP fields, we have to define a quantitative metric to evaluate and compare the priorities of each possible choice. As discussed in the previous section, there are many available criteria and they have a complex interplay with each other. Nevertheless, a simple metric can be devised as guidance for a more educated guess. Since fixing a field implies freezing the pool from which the target list is drawn (according to the requirements for samples P1-P2-P4-P5 specified in the SRD), our approach is 1) to define a prioritization metric at target level, then 2) to define the same metric at the field level by evaluating its integral over the list of P1 targets, which are considered to be those that are most crucial for the success of PLATO. It is a natural first step to review what has been done in the past to solve the same problem by Kepler and TESS teams.

A first proposal of the Kepler Field was centered very close to the Galactic plane, at $b\simeq 5^\circ$ \citep{Borucki2003,Koch2004}, but then revised by increasing its Galactic latitude by ten degrees to mitigate the impact of false positives \citep{Jenkins2005,Batalha2006}. Unfortunately, no detail about this optimization process has ever been published. On the other hand, while the sectors pointed by TESS have a very small margin of adjustment given its fixed sky-scanning law \citep{Ricker2015}, its target-level prioritization scheme has been presented and discussed in great detail by  \citet{Stassun2018,Stassun2019}. The basis of the TESS metric is to compute a quantity proportional to S/N at which a planet of fixed radius, with depth $\delta$ and transit duration $d$, is detected transiting a star of radius $R_\star$ observed by TESS on $N_s$ sectors with an average photometric precision $\sigma$ (mostly dependent on its magnitude in the TESS band). A simple calculation \citep{Stassun2018} based on white noise assumptions returns the following:
\begin{equation}\label{eq:metric_tess}
\Pi_\textrm{TESS} = \frac{S}{N} = \frac{\delta\sqrt{d}}{\sigma / \sqrt{N_s}} \propto \frac{R_\star^{-1.5} \sqrt{N_s}}{\sigma}\quad\textrm{.}
\end{equation}
This quantity is then arbitrarily normalized to lie in the $[0,1]$ range and tuned by manually adding penalizing factors to de-prioritize or exclude specific classes of targets: stars at $b\leq 10^\circ$ (to avoid extremely crowded regions), stars at $\beta\leq 6^\circ$ (not to be observed by the nominal 2-year mission), and stars with unphysical or suspicious physical parameters listed in the TESS Input Catalog (TIC; \citealt{Stassun2019}). 

\begin{figure*}
    \centering
    \includegraphics[width=0.95\columnwidth]{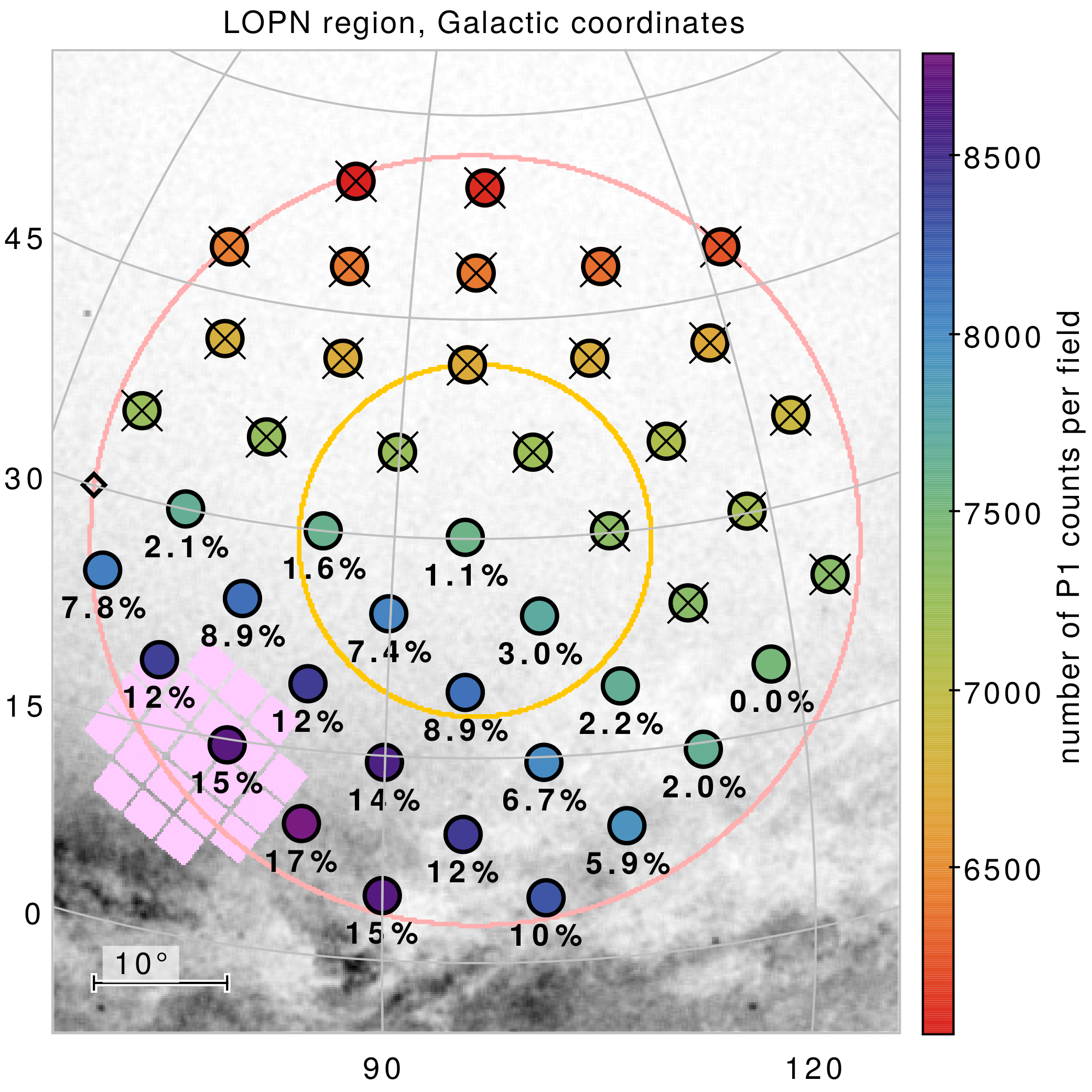}\hspace{4mm}    \includegraphics[width=0.95\columnwidth]{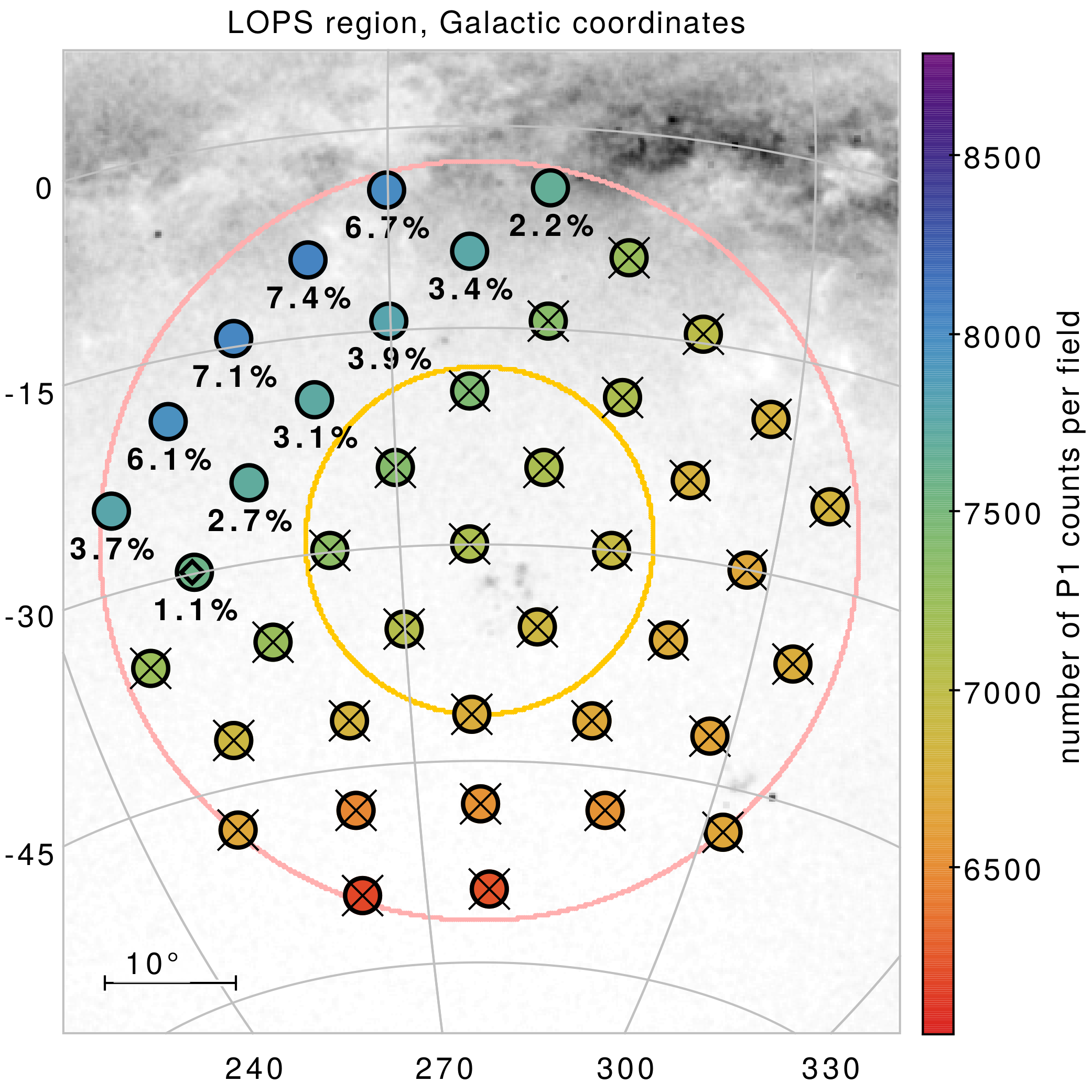}
    \caption{{Grid-based approach for the LOP field selection, step 1} (coarse grid). A coarse grid is defined within the allowed regions at $|\beta| > 63^\circ$ (pink circle) from a HEALPix level-3 scheme in Galactic coordinates. The 86 grid points, corresponding to the centers of 86 possible PF choices, are color-coded according to the number of P1 targets available within each field. The grid points not meeting the science requirement of $>7\,500$  P1 targets per field are marked with a cross, while the 32 surviving fields are labeled with the percentage increase of P1 targets with respect to the minimum requirement. The background gray layer is color-coded according to the areal density of $G<13.5$ stars from Gaia EDR3. The TESS CVZ at $|\beta|\gtrsim78^\circ$ (yellow circle), the Kepler footprint (lilac region), and the centers of the previous provisional fields LOPN0 and LOPS0 (black empty diamond points) are also shown.}
    \label{hpix3}
\end{figure*}

\begin{figure*}
    \centering
    \includegraphics[width=0.95\columnwidth]{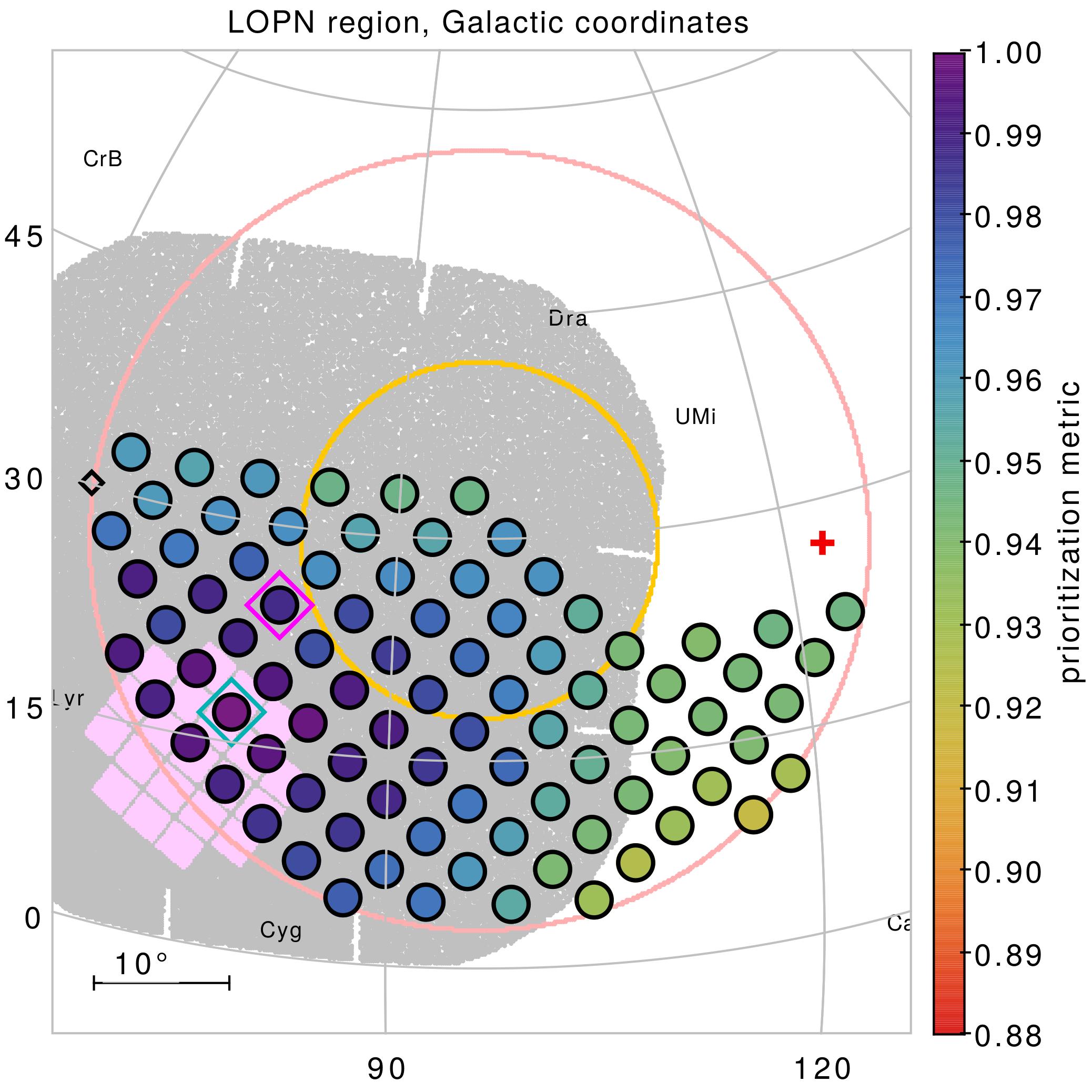}\hspace{4mm}
    \includegraphics[width=0.95\columnwidth]{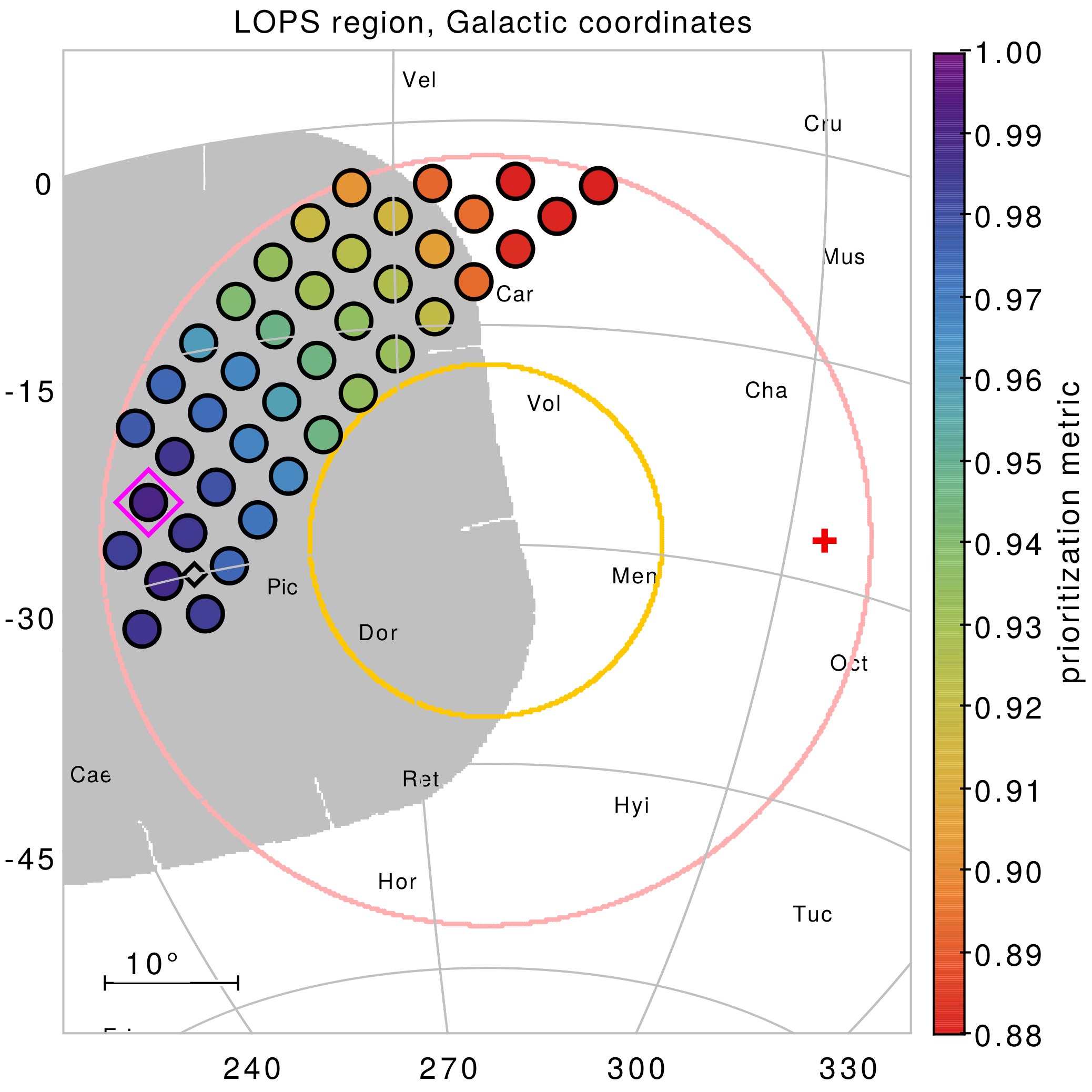}
    \caption{{Grid-based approach for the LOP field selection, step 2} (fine grid). A finer grid is defined within the sky regions compliant with the P1 requirement (see Fig.~\ref{hpix3}), from a HEALPix level-4 scheme in Galactic coordinates. The 128 grid points, corresponding to the centers of 128 P1-compliant PF choices, are color-coded according to the value of the prioritization metric introduced in Section~\ref{metric}, Eq.~\ref{eq:metric}. 
    The LOPN1 and LOPS1 fields newly identified in Section~5 have their footprints plotted as background gray regions and their centers marked with empty magenta points. In the case of LOPN1 (left panel), the final choice does not correspond to the formally highest value of the metric (cyan diamond point; see text for details). The TESS CVZ at $|\beta|\gtrsim78^\circ$ (yellow circle), the Kepler footprint (lilac region), the centers of the previous provisional fields LOPN0 and LOPS0 (black empty diamond points), and the celestial poles (red crosses) are also shown.}
    \label{hpix4}
\end{figure*}

In our case, we could ask ourselves how to develop a metric based on Eq.~\ref{eq:metric_tess}, but that is more suited to the PLATO LOP field selection. First of all, the $\sqrt{N_s}$ factor does not make sense in our case and can be set to one, because by design every LOP target is continuously monitored throughout the nominal duration of the LOP field. Then, as mentioned above, this target-based metric should be summed over the total number of P1 targets within a given field, that is:
\begin{equation}\label{eq:metric0}
\Pi_{\textrm{PLATO}} = \sum_{i\,\in\,\textrm{P1}}\frac{R_{\star,i}^{-1.5}}{\sigma_i}\quad\textrm{,}
\end{equation}
in which $R_{\star}$ can be extracted from the All-Sky PLATO Input Catalog (asPIC; \citealt{Montalto2021}), where all the basic astrophysical parameters for every eligible P1-P2-P4-P5 target in the sky are homogeneously extracted from Gaia~DR2 astrometry and photometry combined with interstellar extinction maps. To give a reliable estimate of the photometric error $\sigma$ (equivalently called noise-to-signal ratio, or NSR\footnote{Here for ``noise'' we mean the total noise, including both the random and systematic components.}), on the other hand, requires a detailed modeling of the optical and electronic performances of PLATO to 1) compute the FOV footprint corresponding to a given PF pointing and 2) estimate $\sigma$ for each asPIC entry, according to its magnitude in the PLATO photometric band\footnote{See \citet{Marchiori2019} for a preliminary definition of the PLATO photometric system.} and its position in the optical FOV and on the CCD detector of each camera. Within the PLATO Consortium, the responsibility of computing the NSR is on the PLATO Performance Team (PPT), which develops and maintains the simulation software.

It is worth investigating how we can enrich the Eq.~\ref{eq:metric0} metric to include other prioritization criteria among those discussed in Section~\ref{criteria}. The most critical missing factor for PLATO, owing to its very large pixel scale ($\sim15"/\textrm{pix}$, with the PSF spreading 90\% of the flux over an area of approximately~$3\times 3$ pixels; \citealt{Rauer2014}) is clearly stellar contamination, both by its photometric and astrophysical components. As for the {photometric contamination}, the PPT simulations already take it into account by estimating the fraction of contaminated flux through a list of neighbor sources provided along with the asPIC. In other words, the impact from photometric contamination is already included in our calculation of $\sigma_i$.

As for the {astrophysical contamination}, which is mostly due to blended, detached eclipsing binaries mimicking a planetary transit, a more sophisticated approach is required. Focusing on the specific characteristics of PLATO, including its pixel scale and PSF size, Bray et al.~(in prep.) exploited a binary population model to estimate the false positive ratio (FPR) (defined as the ratio between FPs and the total number of detections) as a function of the line of sight, and for different bins of planetary radius. As expected, the Galactic latitude $b$ is the key variable in determining the FPR, the dependence on longitude $l$ being very weak and not even statistically significant for most bins. After neglecting the $l$ dependence, we fit the parameters of an exponential law to the $(|b|, \texttt{FPR})$ relation found for terrestrial-sized planets by Bray et al.:
\begin{equation}\label{eq:bray}
\texttt{FPR}(b)=0.211\cdot 10^{-0.035\,\cdot\, |b|} \textrm{ ,}\qquad 0<\log\left (\frac{R_p}{R_\oplus}\right )<0.2 \textrm{ ,}
\end{equation}
with the planetary radius range corresponding to $1<R_p/R_\oplus\lesssim 1.58$ in linear units. As a simplifying step, we can assume that the ``value'' of a given target is inversely proportional to the additional follow-up time wasted in rejecting the FPs. We can therefore modify our metric by introducing a factor $(1-\texttt{FPR})$ to give, as an example, 
 half weight to a P1 target with a 50\% FPR, and so on. The field-based metric is then as follows:
\begin{equation}\label{eq:metric}
\Pi_{\textrm{PLATO}}' = \sum_{i\,\in\,\textrm{P1}}\left[\frac{R_{\star,i}^{-1.5}}{\sigma_i}\times \left ( {1-\texttt{FPR}(b)_i} \right ) \right]\textrm{ .}
\end{equation}

In the next section, we evaluate and compare the $\Pi_{\textrm{PLATO}}'$ quantity for a set of different LOP choices as guidance in the field selection process. Adding too many parameters at once in such a complex task would be confusing, and it would create an issue about how these additional terms are arbitrarily weighted with each other. Rather, we discuss the other prioritization criteria {a posteriori}, after having identified the sky regions where the expression in Eq.~\ref{eq:metric} is maximized.

\section{The grid-based approach for field selection}\label{grid}

Simulating the geometry and content of a given PF is a computationally intensive task, especially when it comes to the noise calculation. To speed up the process, rather than maximizing $\Pi_{\textrm{PLATO}}'$ as a continuous function, we implemented a two-staged, grid-based approach. We first defined:
\begin{enumerate}
    \item a {coarse grid} to identify the sky regions where the formal requirements (Sec.~\ref{requirements}) are met, then
    \item  a {finer grid} where $\Pi_{\textrm{PLATO}}'$ (Eq.~\ref{eq:metric}) could be individually computed and the additional scientific criteria assessed (Sec.~\ref{criteria}).
\end{enumerate}

\subsection{The ``coarse'' grid: Checking the requirements}

The coarse grid was built starting from an all-sky HEALPix level-3 partition \citep{Wallace2005}, indexed with the RING scheme in Galactic coordinates, made of 768 points with an average spacing of $7.3^\circ$. A subset of 86 pointings allowed by the $|\beta| > 63^\circ$ constraint was then selected (plotted as big circles in both panels of Fig.~\ref{hpix3}). After having ingested asPIC~1.1 as input catalog, for each field pointing the PPT simulated the photometric performances of PLATO with the most updated instrumental model, delivering back to us the corresponding 86 stellar catalogs inclusive of the expected NSR $\sigma_i$ for the individual targets (and hence their membership flag to the P1-P2-P4-P5 samples). The simulations generating the NSR estimates were carried out with the PINE software (B\"orner et al., in prep.). As explained in Section~\ref{metric}, this simulation takes into account the additional noise from background sources (i.e., the photometric contamination). 
It is then straightforward to extract the P1-P2-P4-P5 counts for each field and check their compliance with the SRD requirements. Such requirements are formally defined as the sum over two LOP fields, thus in a broad sense we refer to a single field as ``compliant'' when it reaches half of the minimum counts specified on the SRD.
As a result, only 32 grid points (21 in the northern Ecliptic cap and 11 in the southern one) meet both the P1 and P5 requirements, with the former being the most stringent one (the non-compliant fields are struck with a black cross; Fig.~\ref{hpix3}). The P2 and P4 requirements are always satisfied, over all the 86 grids points. 

As a general trend, the P1 counts increase toward the Galactic plane (by up to $\sim\!\!17\%$ within the compliant region). The northern region (Fig.~\ref{hpix3}, left panel) is on average richer in P1 targets than the southern one (right panel); not only is the compliant area larger for the former, but also the count gradient as function of $|b|$ is stronger. The P5 counts (not plotted) show an even stronger gradient (up to $\sim\!\!40\%$), which is not surprising since a fainter magnitude limit translates into a longer distance probed into the Galactic disk.
 On the other hand, the count dependence on Galactic longitude is much more complex: For the P1 sample, it is mostly negligible in the LOPN region, but it is not in the LOPS region (where its impact is about $\sim\!\!20\%$ along Galactic parallels), while for the P5 sample the trend is completely reversed. The asymmetric behavior between the north and south caps may look puzzling at first sight, but actually it is not unexpected due to the very different distribution of Galactic dust along these lines of sight (\citealt{Lallement2019}; see Section~\ref{content} for more details). We emphasize that the impact of interstellar extinction on targets usable by PLATO is the result of a complex interplay because its ``local'' component  decreases the number of available targets by increasing their apparent magnitude and noise level, while, at larger distances, it mitigates both photometric and astrophysical contamination by masking background sources. As our stellar samples are magnitude-limited, this ``local'' threshold is of course dependent on that limit, being $d\lesssim 300$~pc for the P1 sample (Fig.~\ref{fig:p1content1}, lower panels). On top of this, the faint tail of the P1 and most of P5 sample probe even larger distances (up to $d\simeq 1000$~pc), where the spiral structure of the Milky Way starts to be discernible \citep{Miyachi2019,Poggio2021} and the basic assumption of spatial homogeneity breaks. 

A closer look at the stellar content of the compliant fields reveals the following: 1) the gain of P1 targets at lower Galactic latitudes is mostly due to the increasing fraction of F-type main-sequence and slightly evolved stars (and, to a minor extent, to K-type subgiants), while the sky density of G and K dwarfs is close to be isotropic; and 2) the actual impact of photometric contamination on the P1 sample is mostly negligible even close to the Galactic plane, where the average increase in photometric noise due to it is at most a few percent. In other words, while there is some numerical gain of targets in moving the LOP fields closer to the disk, this gain is due to the less valuable stars in terms of scientific output because F dwarfs are larger and more massive than the Sun and with a higher rotational velocity, posing challenges to both transit detection and RV follow-up. We come back to this in Section~\ref{content}.

\subsection{The ``fine'' grid: Optimizing the metric}

The fine grid was built starting from an all-sky HEALPix level-4 partition, made of 3072 points with an average spacing of $3.7^\circ$. Only points within the $|\beta| > 63^\circ$ allowed region and broadly overlapping with the ``compliant'' region found in the previous section were retained (i.e., non-crossed circles of Fig.~\ref{hpix3}). The 128 survived grid points, 87 in the north and 41 in the south, are plotted in Fig.~\ref{hpix4}. Again, for each of the 128 fields, a stellar catalog of all the observed targets including the calculated NSR $\sigma_i$ has been provided by the  PPT with the same procedure described above. Finally, for each field, we computed the $\Pi_{\textrm{PLATO}}'$ metric as defined in Eq.~\ref{eq:metric}; the output value is color-coded with the same scale in both panels of Fig.~\ref{hpix4}. It is worth emphasizing that this dimensionless metric has only a relative meaning in our optimization process, while its absolute normalization, unlike other quantities such as the target counts, is arbitrary and not informative. For this reason, the maximum value of $\Pi_{\textrm{PLATO}}'$ over the whole LOPS+LOPN grid (i.e., 121.16) has been normalized to one in the color scale of both panels of Fig.~\ref{hpix4}.

\section{Identification of the provisional LOP fields}\label{identification}

A quick look at the smooth spatial distribution of $\Pi_{\textrm{PLATO}}'$ (Fig.~\ref{hpix4}) reveals two well-localized ``sweet spots'' (in the LOPS and LOPN regions, respectively) where our prioritization metric reaches a global maximum. In the following we discuss the identification of the two LOP fields separately.

\subsection{The southern LOP field: LOPS1}

\begin{figure*}[h!]
    \centering
    \includegraphics[width=0.9\columnwidth]{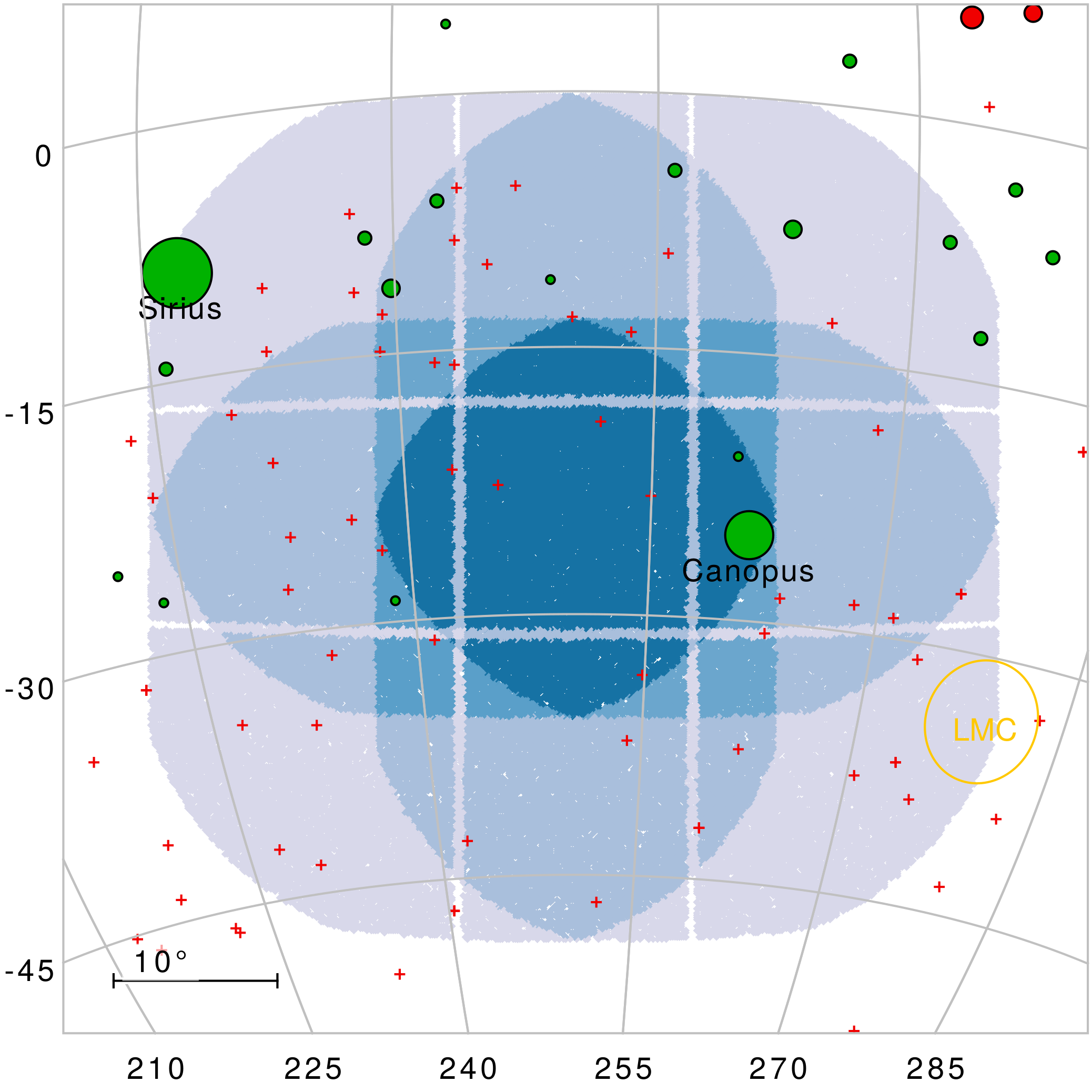}\hspace{6mm}
    \includegraphics[width=0.9\columnwidth]{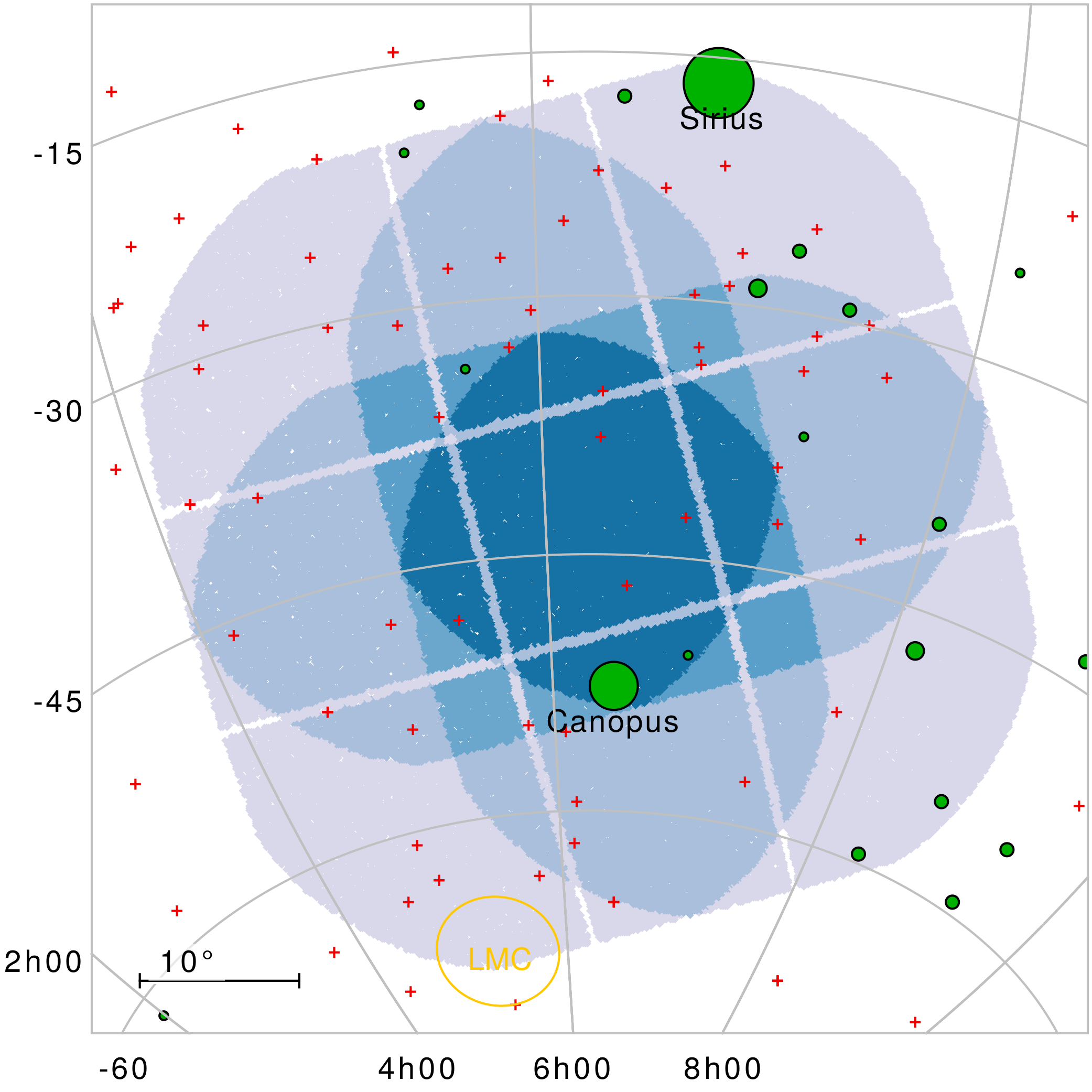} \\ \vspace{6mm}
    \caption{{Location of the provisional South PLATO LOP Field (LOPS1)} in a Galactic (left panel) and equatorial (right panel) projection. The PF is color-coded in blue shades according to the number of co-pointing ``normal'' cameras, from six (light blue) to 24 (dark blue). Stars brighter than $V=3$ are plotted with green circles; their area is proportional to their $V$-band flux. Known exoplanets are plotted with red crosses; none of them reach the threshold of 50 bibliographic references. The position of the Large Magellanic cloud is over-plotted with a yellow ellipse.}
    \label{fig:LOPS1planets}
\end{figure*}

Within the LOPS region, the metric has its maximum value at the grid point labeled \#2188 in the HEALPix RING scheme, located at Galactic coordinates $l\simeq 250^\circ.3$, $b\simeq -24^\circ.6$ (at the boundary between Columba and Puppis) and marked with a magenta diamond point in the right panel of Fig.~\ref{hpix4}. A PF centered at this line of sight (gray outline, same plot; also Fig.~\ref{fig:LOPS1planets}) would nearly graze the Galactic plane in the six-telescope region, with its northernmost border reaching $b\simeq-0^\circ.25$; the whole Col, Pic, Cae, and Dor constellations and parts of Car, Pup, CMa, Lep, Ret, Hor, and Eri are covered by this PLATO field. The central FOV region covered by 24 telescopes, where the highest photometric precision is achieved, lies in the $-35^\circ\lesssim b\lesssim-15^\circ$ range. This is also where most of the P1 targets are expected to be due to the 50~ppm in one hour noise requirement.
From now on, we refer to this pointing as the new provisional location of the southern LOP field, called LOPS1; it lies only a few degrees from the previous LOPS0 (black diamond point in the plot), while having a higher metric value. The coordinates of its center are shown in Table~\ref{table:coord} in different reference frames, together with the number of corresponding P1-P2-P4-P5 counts. As from construction, all of them fulfill the formal SRD requirements. A discussion about the LOPS1 content and its astrophysical properties is given in Section~\ref{content}.

\begin{figure*}[h!]
    \centering
    \includegraphics[width=0.9\columnwidth]{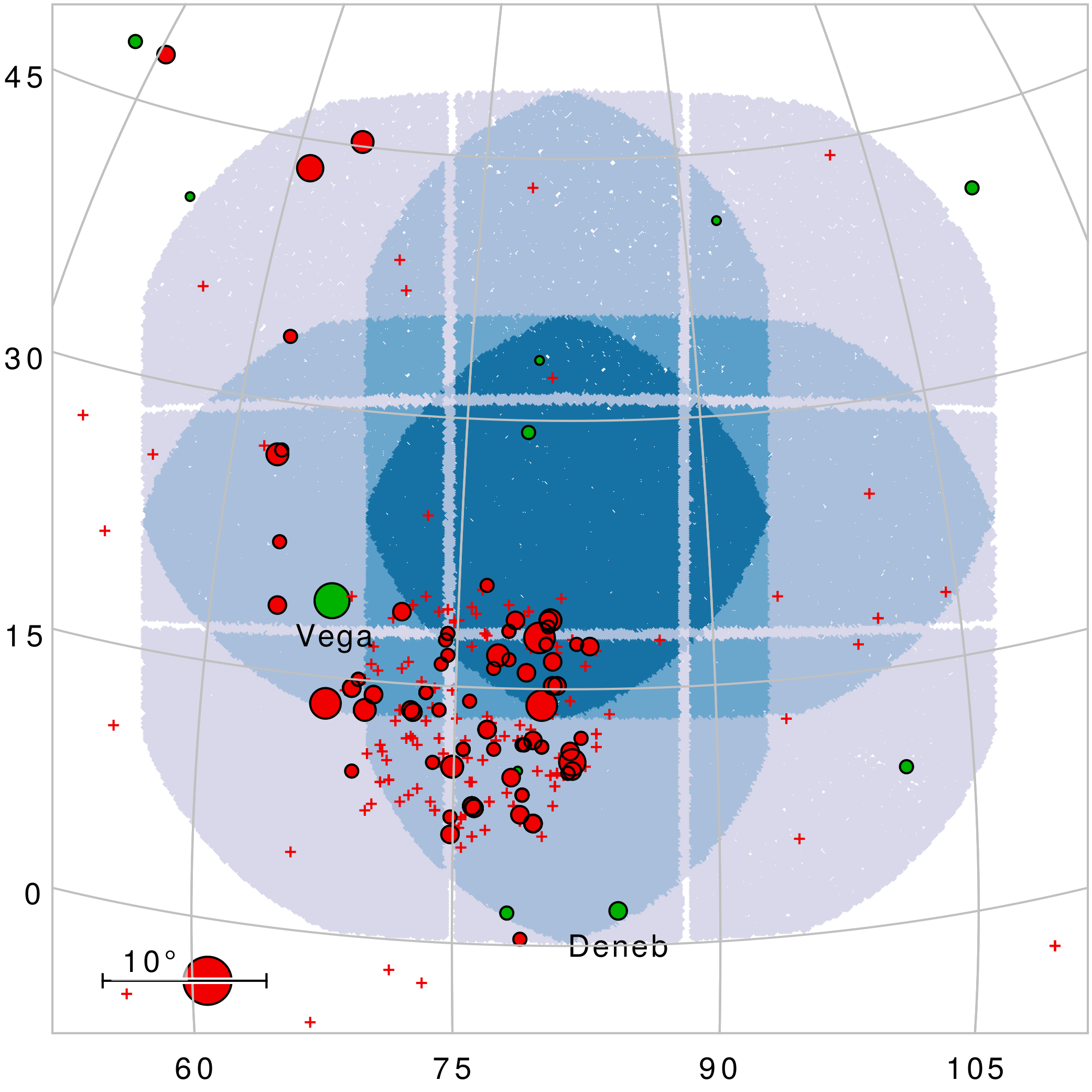}\hspace{6mm}
    \includegraphics[width=0.9\columnwidth]{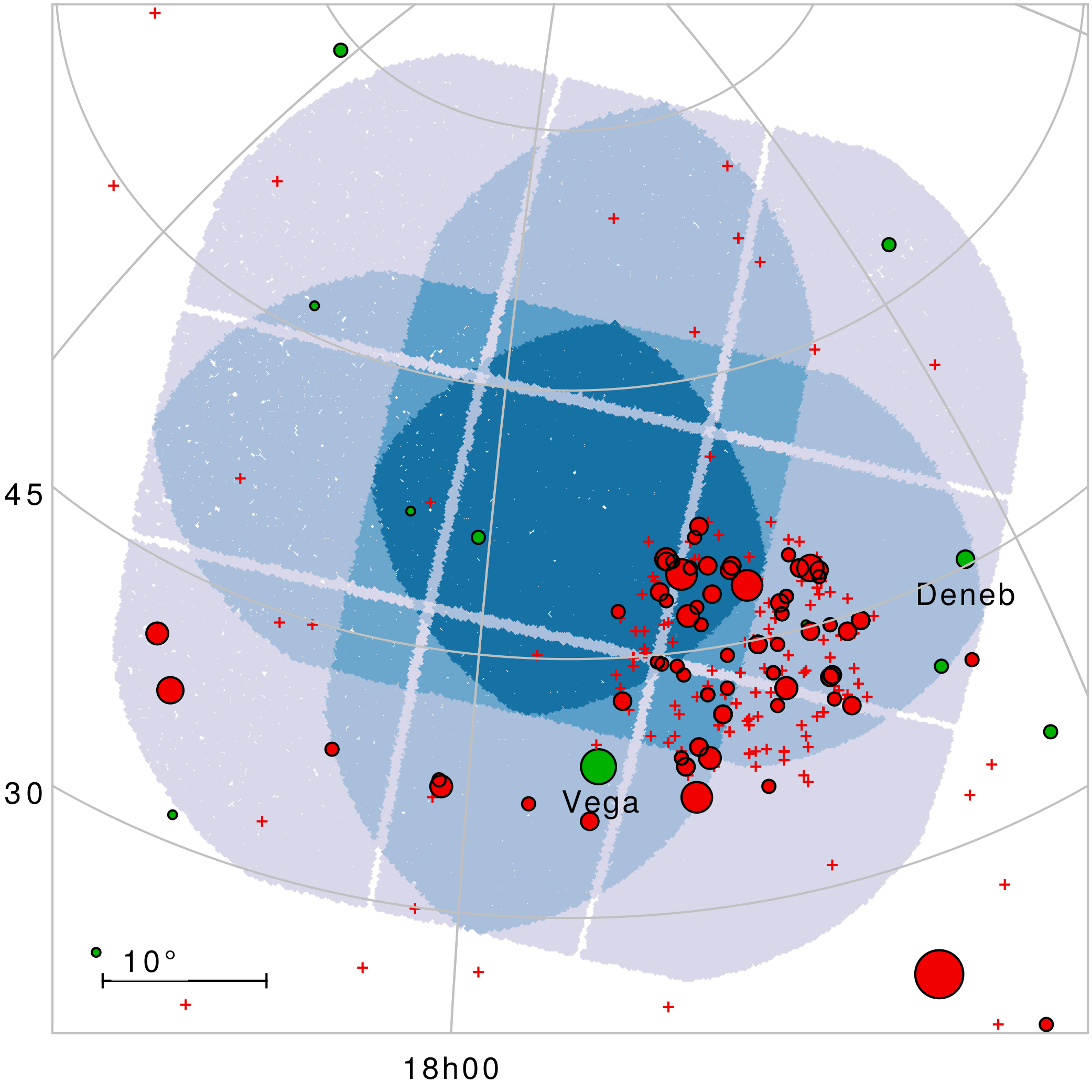} \\ \vspace{6mm}
    \caption{{Location of the provisional North PLATO LOP Field (LOPN1)} in a Galactic (left panel) and equatorial (right panel) projection. The PF is color-coded in blue shades according to the number of co-pointing ``normal'' cameras, from six (light blue) to 24 (dark blue). Stars brighter than $V=3$ are plotted with green circles; their area is proportional to their $V$-band flux. Known exoplanets are plotted with red crosses; those with more than 50 bibliographic references are over-plotted with red circles and their area is proportional to the number of references.}
    \label{fig:LOPN1planets}
\end{figure*}

\subsection{The northern LOP field: LOPN1}

Evaluating $\Pi_{\textrm{PLATO}}'$ for the northern field requires more reasoning. The region where the metric has a global maximum is larger and flatter with respect to the LOPS1 choice, encompassing a vast area approximately centered on the north rim of the Kepler Field (Fig.~\ref{hpix4}, left plot). Some of those points lie at quite low Galactic latitudes, $10^\circ\lesssim b\lesssim 20^\circ$, implying that the corresponding PFs cross the Galactic plane and include a significant fraction of the FOV on extremely crowded stellar fields. Selecting the grid point with the nominally highest value of $\Pi_{\textrm{PLATO}}'$, at $l\simeq 78^\circ.7$, $b\simeq 16^\circ.9$ (HEALPix label: \#1070; cyan diamond point on Fig.~\ref{hpix4}), for instance, would imply having $\sim 30\%$ of the P1 targets and $\sim 42\%$ of the P5 targets at $|b|<10$, respectively. While our metric takes both photometric and astrophysical contamination into account to some extent, choosing such a low-latitude pointing could make additional issues arise. Those include, not only the impact from less common blending scenarios not considered in the Bray et al.~analysis and that are much more difficult to model, but also an increased difficulty in characterizing the target and its contaminants at the validation and confirmation stage, when information from external catalogs and databases have to be merged into a global modeling. With this in mind, it makes sense to ponder what the actual scientific gain of such a choice would be with respect to a choice at $|b|\simeq 25^\circ$, that is,~as for LOPS1. Focusing on the P1 sample, the gain in terms of counts is just $\sim 7\%$. A careful comparison on how the astrophysical parameters are distributed reveals that the number of main-sequence G and K stars in that sample is mostly independent of the field choice, as these (nearby) targets are isotropically distributed on the sky. Rather, the gain in counts is almost entirely due to F stars at $d\gtrsim 200$~pc, and, to a lesser extent, to distant cool subgiants. When it comes to the core objectives of the PLATO mission, the latter two components are less scientifically valuable with respect to GK dwarfs, both from a detection point of view (larger radii and hence smaller transit depths at a given planetary radius) and at the follow-up stage (larger masses and higher rotational velocity making ultra-high-precision RVs more difficult and expensive in terms of observing time).

As a much safer alternative, we can select as LOPN1 (provisional North LOP field) the best pointing located at $|b|>20^\circ$, identified by the HEALPix index \#878 (marked by a magenta diamond point in the left panel of Fig.~\ref{hpix4}; also Fig.~\ref{fig:LOPN1planets}). The value of our prioritization metric differs by only $1\%$ with respect to the \#1070 pointing. LOPN1 is located at Galactic coordinates $l\simeq 81^\circ.6$, $b\simeq 24^\circ.6$ (in Draco) and marked with a magenta diamond point in the right panel of Fig.~\ref{hpix4}, with its footprint plotted as a gray area. A large fraction of Lyr, Dra, Cyg, and Her constellations and parts of Cep and UMi are covered by LOPN1. The coordinates of its center are shown in Table~\ref{table:coord} in different reference frames, together with the number of corresponding P1-P2-P4-P5 counts. As for LOPS1, all of them fulfill the formal SRD requirements. A discussion about the LOPN1 astrophysical properties is given in the following section.

\section{Characterization of the provisional LOP fields}\label{content}

In the previous sections, we have identified LOPS1 and LOPN1 as the best provisional PFs to be observed during the LOP phase, and demonstrated that they both meet the mission formal requirements and optimize our scientific prioritization metric. Now we can take a closer look at their content and its astrophysical properties. The full footprints of LOPS1 and LOPN1, which are color-coded to show the coverage of different regions by 24-18-12-6 cameras, are shown in Fig.~\ref{fig:LOPN1planets} and Fig.~\ref{fig:LOPS1planets} in Galactic and equatorial coordinates. As mentioned above, both fields, centered at $|b|\simeq 25 ^\circ$, span a very wide range of Galactic latitudes, from the Galactic plane up to nearly $|b|\simeq 50^\circ$. In terms of declination $\delta$, both fields avoid the celestial poles, most targets being in the $30^\circ\lesssim|\delta|\lesssim 60^\circ$ band, that is, they are fully within the range which maximizes the efficiency and schedulability of the ground-based follow-up (Sec.~\ref{fu}; Fig.~\ref{fig:fu}). A non-negligible fraction of LOPS1, in particular, is monitored by 6-12 cameras at $-30^\circ\lesssim\delta\lesssim -15^\circ$ (Fig.~\ref{fig:LOPS1planets}; right plot), that is limitedly observable even by facilities in the northern hemisphere. 

\subsection{Bright stars}\label{content:brightstars}

A few extremely bright stars fall on silicon within our provisional fields (green circles on Fig.~\ref{fig:LOPN1planets} and \ref{fig:LOPS1planets}), most notably Vega = $\alpha$~Lyr and Deneb = $\alpha$~Cyg in LOPN1, and Sirius = $\alpha$~CMa and Canopus = $\alpha$~Car in LOPS1. This is unavoidable when dealing with such a wide FOV. Indeed, a formal requirement from SRD states that ``stars brighter than the maximum dynamic range shall not impede operation of the instruments, except by modifying the number of useful pixels''. During the future fine-tuning stage of the LOP fields, it will be investigated whether it is advisable to slightly adjust the field position in order to move some of those ultra-bright stars out of the FOV (as it is feasible for Sirius and Deneb) or within the CCD gaps of at least a subset of cameras (e.g.,~for Vega). The impact of stray light from these extremely bright stars should be evaluated.

\subsection{Known exoplanets}\label{content:exoplanets}

It is of scientific interest to find out which already known (confirmed) transiting exoplanets will be imaged during the LOP phase, as all of them will be added in the target list regardless of their formal inclusion in the P1-P2-P4-P5 samples. We started from the latest online version of the \texttt{TEPCAT} catalog \citep{Southworth2011} and cross-matched it with Simbad \citep{Wenger2000} to get the total number of bibliographic items referring to that target for each entry, as a very rough proxy for  its scientific ``popularity''. All of those planets (51 in the LOPS1 footprint and 192 in the LOPN1 one) are plotted as red symbols in Fig.~\ref{fig:LOPN1planets} and \ref{fig:LOPS1planets}; an arbitrary threshold of 50 publications has been highlighted by a different plotting style to make the most ``popular'' targets visually stand out. Not surprisingly, most of the known LOPN1 planets are actually Kepler objects since the Kepler Field is fully overlapped with it.  
Other non-Kepler, well-studied planetary systems are TrES-1, TrES-4, WASP-3, and HAT-P-5 (in the 12-camera regions) as well as KELT-9, HAT-P-14, HD149026, and HAT-P-2 (in the six-camera region); the recently discovered ultra-short-period TOI-1444, to be monitored with 12 telescopes is worth mentioning for its small radius (1.44~$R_\oplus$; \citealt{Dai2021}). On the LOPS1, none of the known planets cross the 50-publication threshold. Nevertheless, at least four confirmed systems host planets smaller than 2~$R_\oplus$, all of them were discovered by TESS: TOI-540 \citep{Ment2021}, LHS~1815 \citep{Gan2020}, TOI-700 \citep{Gilbert2020}, and TOI-270 \citep{Gunther2019}. 

\begin{table}\centering
  \caption{Properties of the provisional LOP fields LOPS1 and LOPN1 (Fig.~\ref{grid}).}
\begin{tabular}{c|rr|l}
\hline
 field & LOPS1 & LOPN1 & notes \\ \hline \hline
 HEALPix & \#2188 & \#0878 \rule{0pt}{16pt} & level $k=4$, \\
 index & & & RING scheme \\
 $\alpha$ [deg]& 93.49134 & 277.18023 & ICRS \rule{0pt}{16pt}\\ 
 $\alpha$ [hms]& 06:13:57.9 & 18:28:43.2 & ICRS\\ 
 $\delta$ [deg]& $-$42.93544 & 52.85952 & ICRS\\
 $\delta$ [hms]& $-$42:56:08 & 52:51:34 & ICRS\\
 $l$ [deg] & 250.31250 & 81.56250 & IAU 1958\rule{0pt}{16pt}\\
 $b$ [deg] & $-$24.62432 & 24.62432 & IAU 1958\\
 $\lambda$ [deg] & 96.36781 & 287.98162 & Ecliptic \rule{0pt}{16pt}\\
 $\beta$ [deg] & $-$66.29759 & 75.85041 & Ecliptic \\
 P1 targets & 7\,806 & 8\,190 & Req.~7\,500 \rule{0pt}{16pt} \\ 
  P2 targets & 699 & 705 & Req.~500\\ 
P4 targets & 17\,115 & 16\,833 & Req.~2\,500\\
P5 targets & 154\,639 & 158\,806 & Req.~122\,500 \\ \hline
\end{tabular}
\tablefoot{The listed quantities are: HEALPix indexing, coordinates for the field centers in equatorial, Galactic, and ecliptic reference frames, and the number of targets falling into the P1-P2-P4-P5 categories. The P1-P2-P4-P5 requirements are formally specified for the sum of the LOP fields; the quantity reported in the ``notes'' column is just half of that value (See Section~\ref{criteria} for more details).}
\label{table:coord}
\end{table}

\begin{figure*}[p!]
    \centering
    \includegraphics[width=0.78\columnwidth]{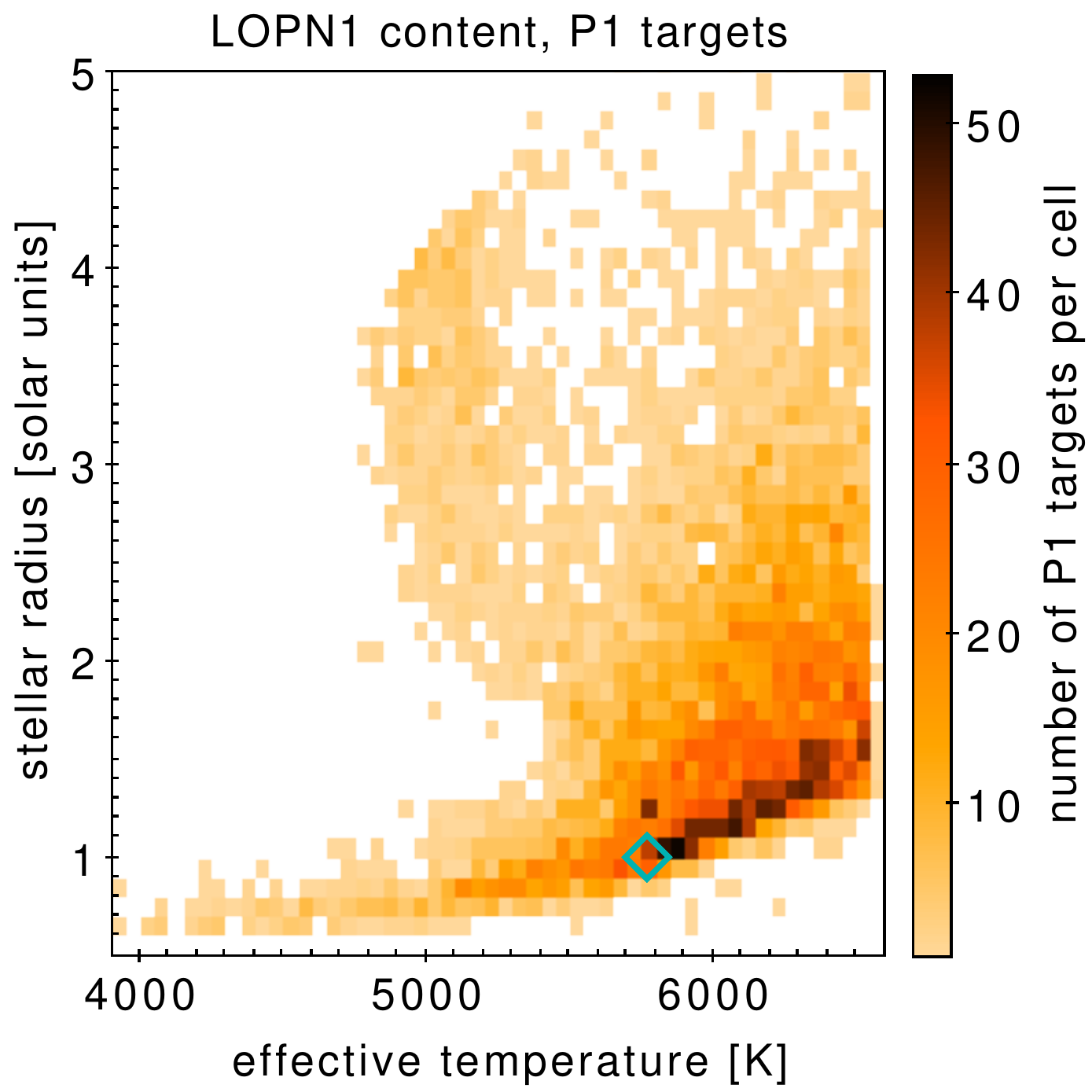} \hspace{1cm}
    \includegraphics[width=0.78\columnwidth]{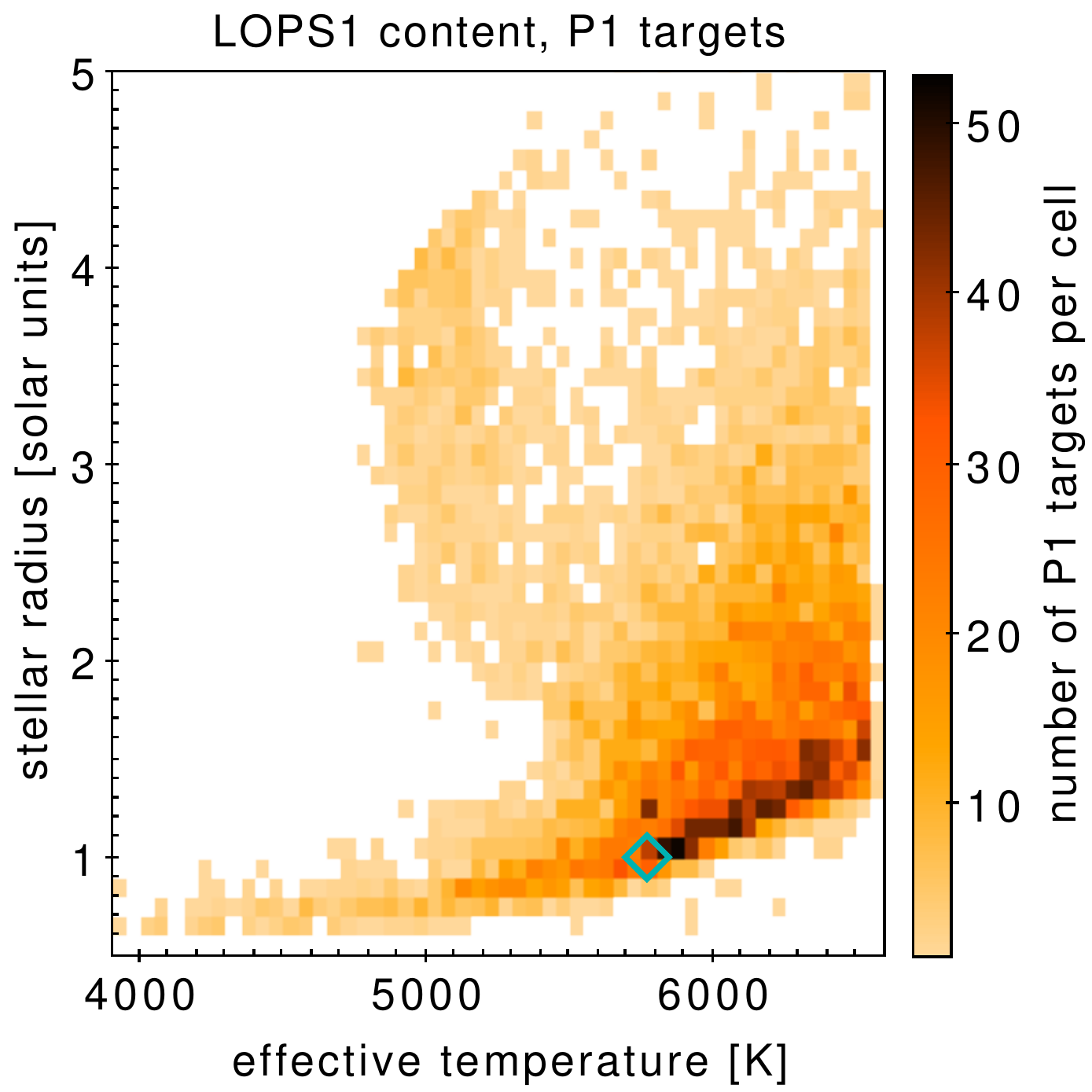} \\ \vspace{0.6cm}
    \includegraphics[width=0.78\columnwidth]{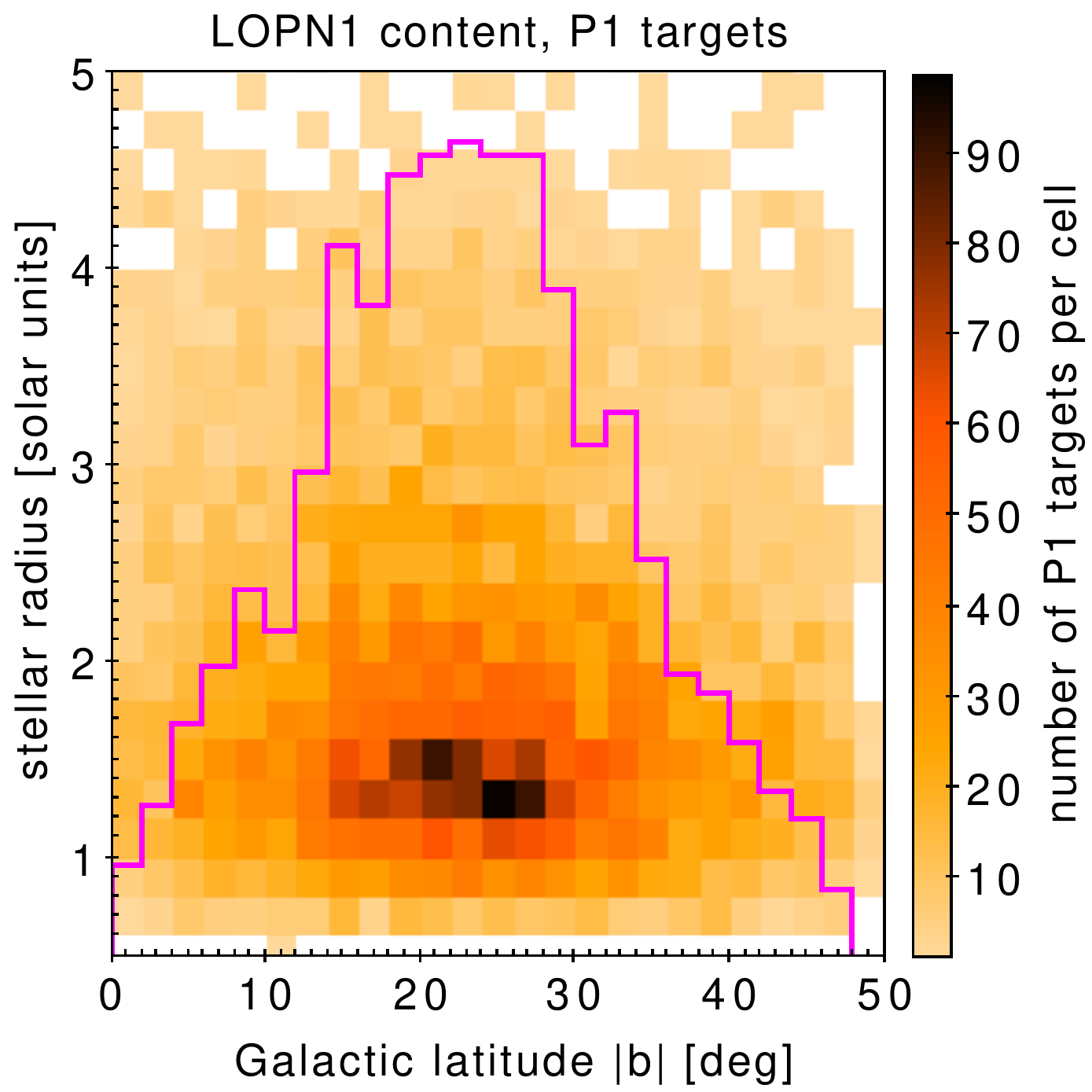} \hspace{1cm}
    \includegraphics[width=0.78\columnwidth]{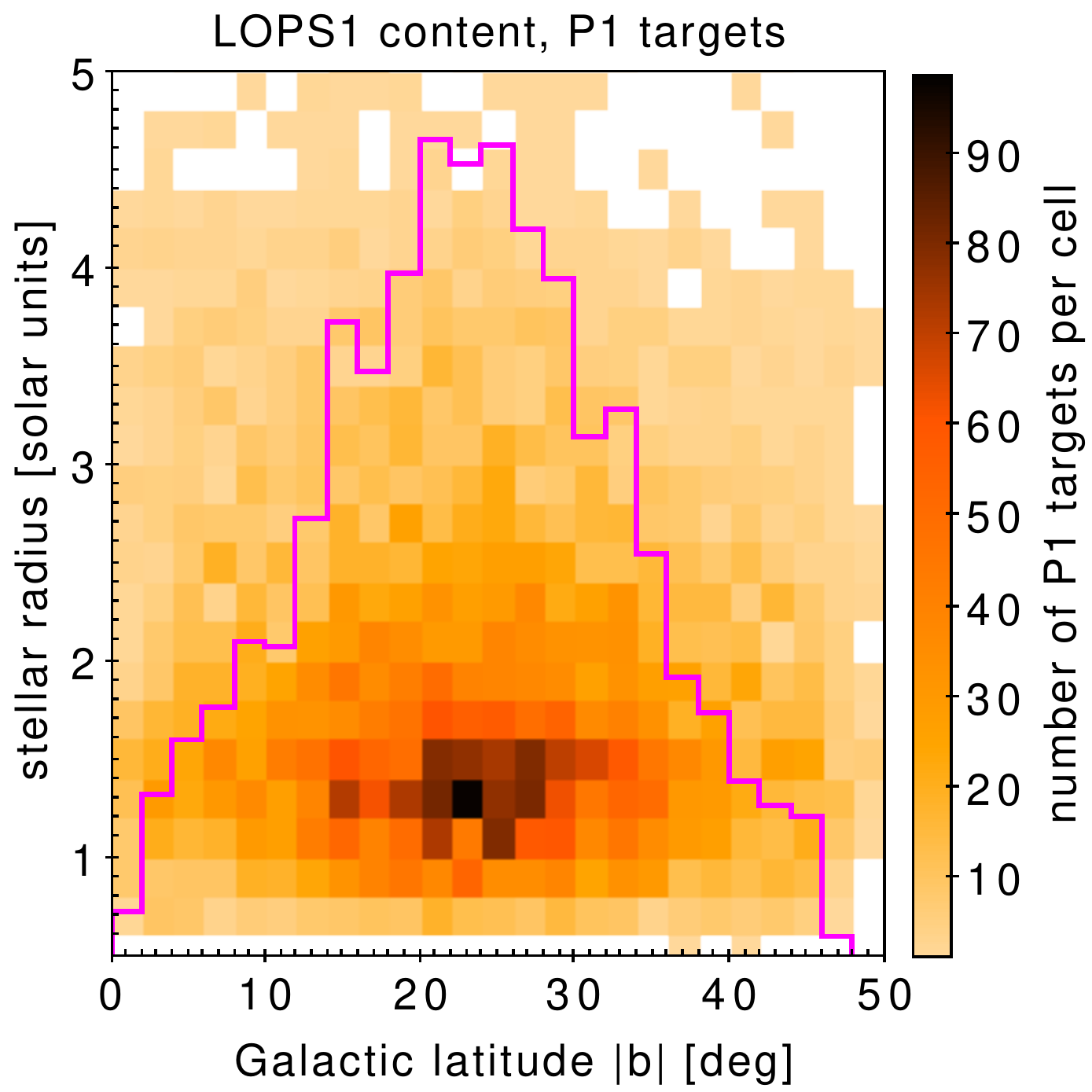} \\ \vspace{0.6cm}
    \includegraphics[width=0.78\columnwidth]{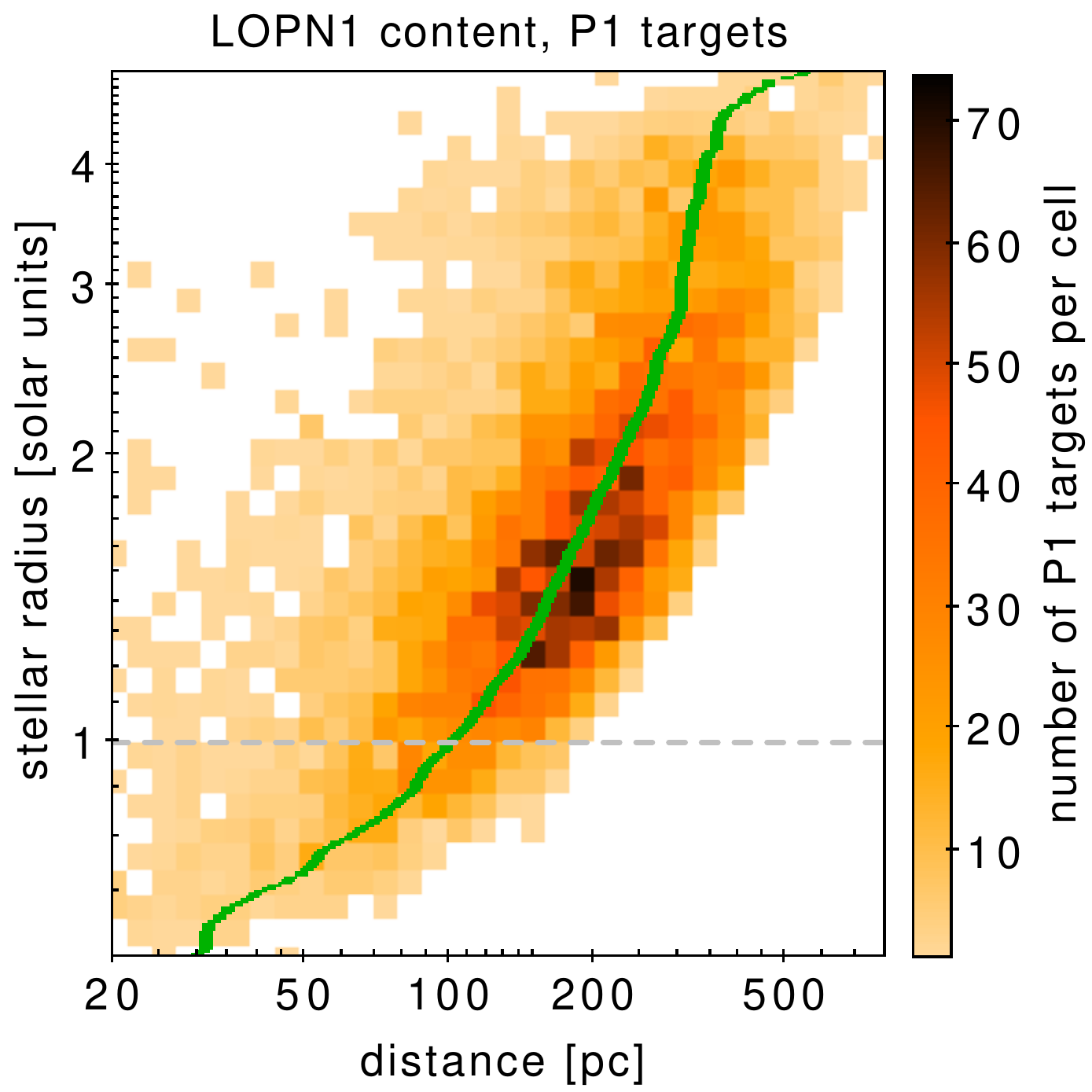} \hspace{1cm}
    \includegraphics[width=0.78\columnwidth]{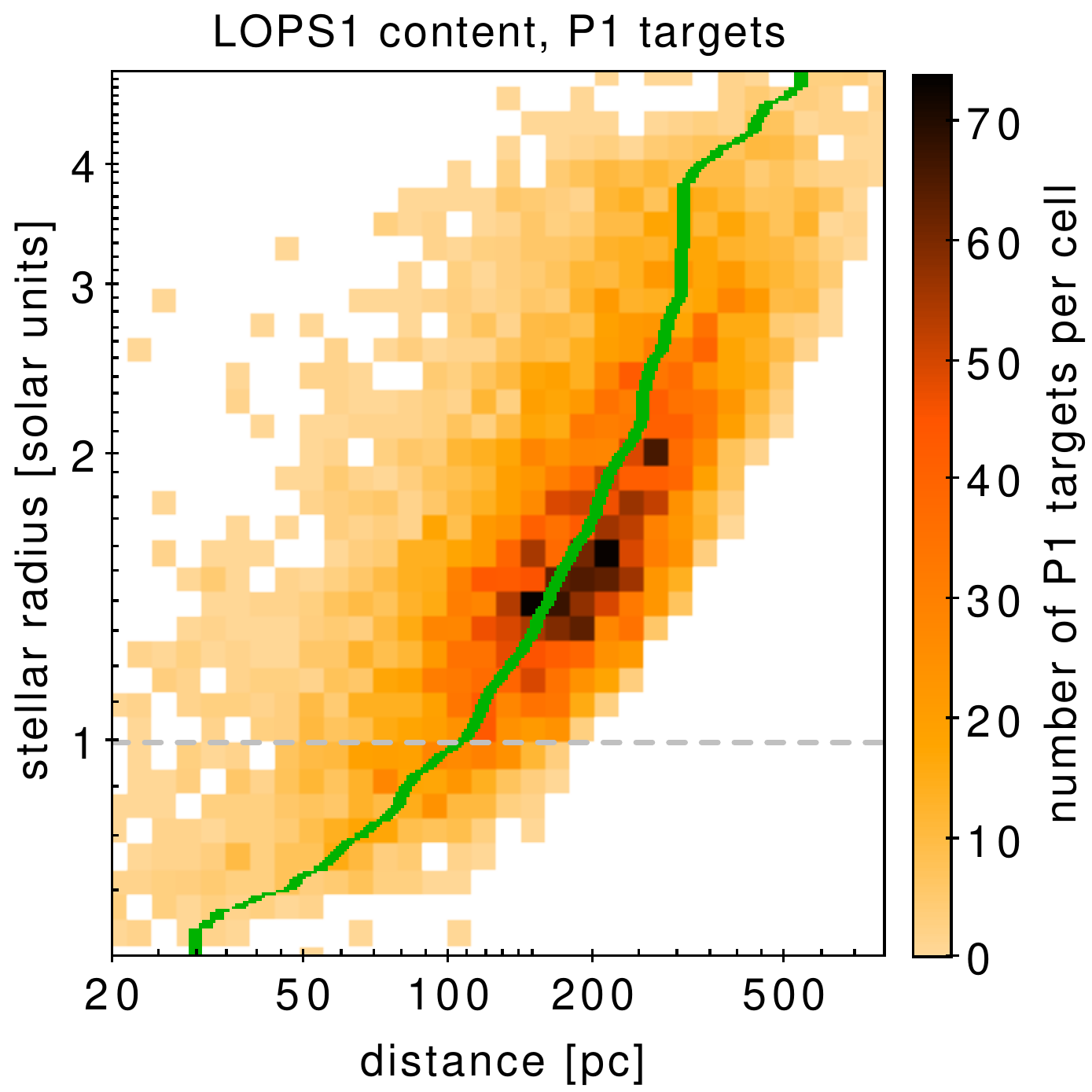}
    \caption{{Astrophysical parameters of the P1 targets} within the LOPN1 (left panels) and LOPS1 (right panels) provisional fields. \emph{Top panels:} Number density of P1 targets as a function of effective temperature and stellar radius. The high-density ridge corresponds to the main sequence, where the Solar parameters (1~$R_\odot$, 5778~K) are marked with a cyan diamond point as reference. \emph{Middle panels:} Number density of P1 targets as a function of Galactic latitude $|b|$ and stellar radius. A histogram of $b$ is overplotted with a magenta line and arbitrary normalization.  \emph{Bottom panels:} Number density of P1 targets as a function of distance and stellar radius. The median distance at fixed solar radius is plotted with a green line. }
    \label{fig:p1content1}
\end{figure*}

\begin{figure*}\centering
    \includegraphics[width=0.8\columnwidth]{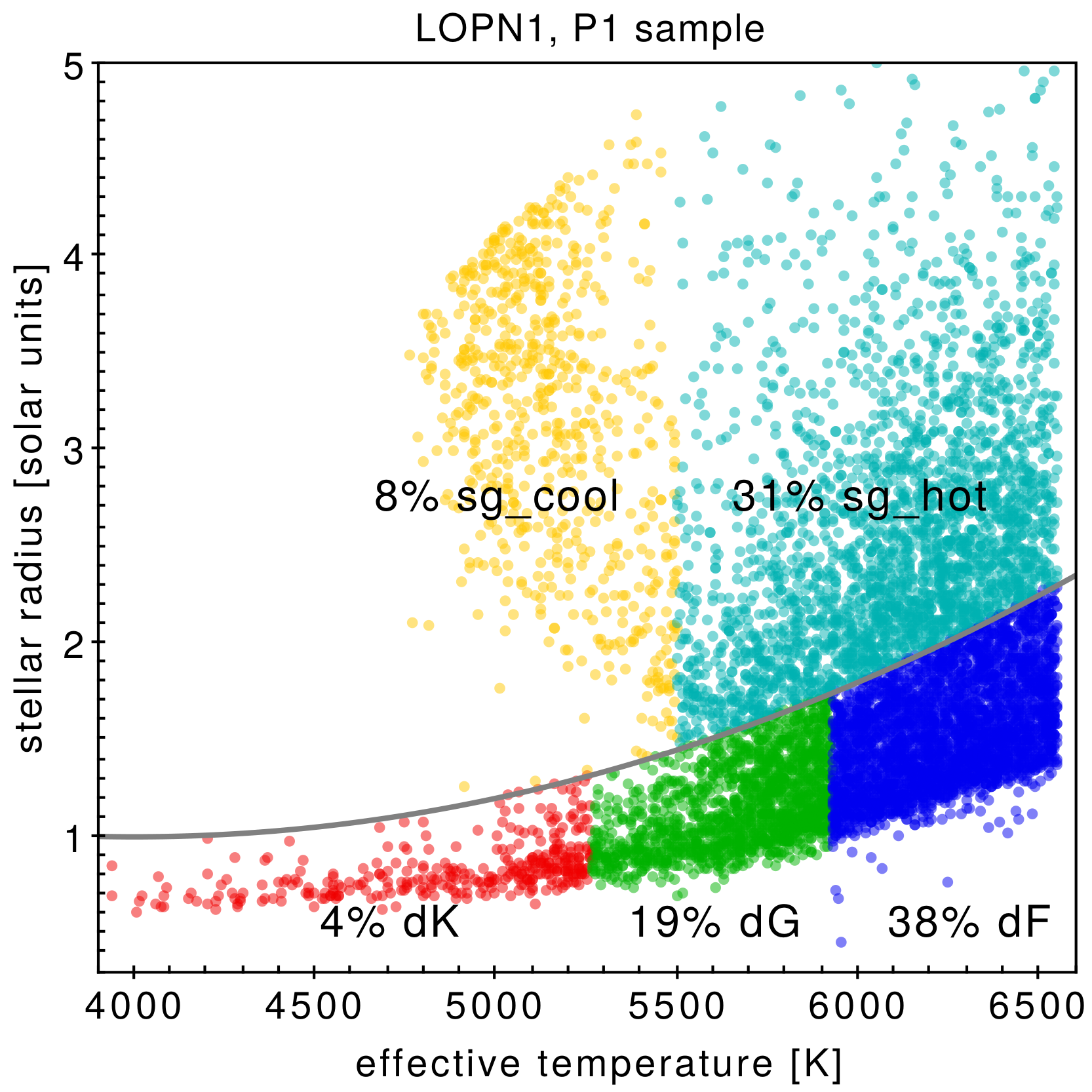} \hspace{1cm}
    \includegraphics[width=0.8\columnwidth]{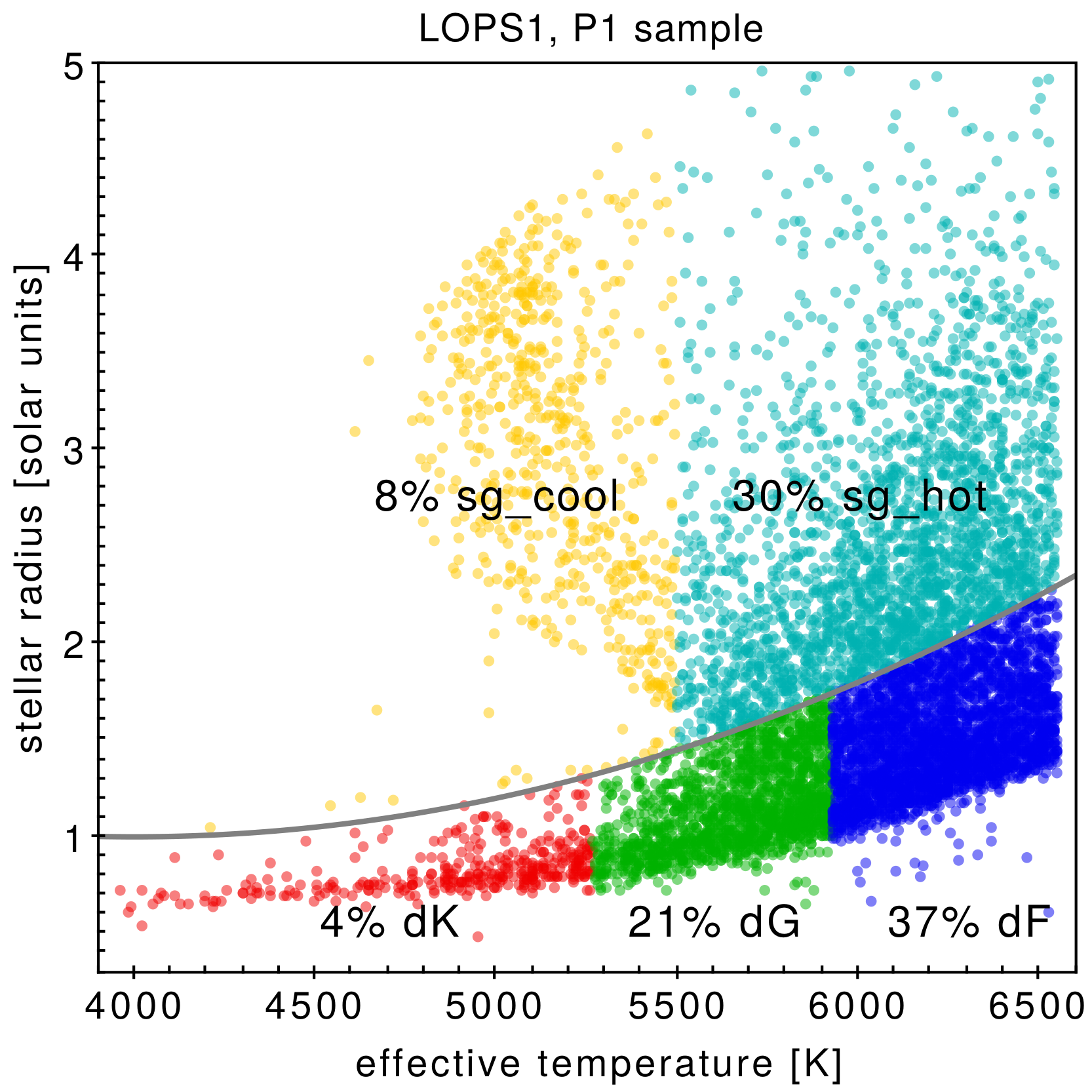} \\ \vspace{0.4cm}
    \includegraphics[width=0.8\columnwidth]{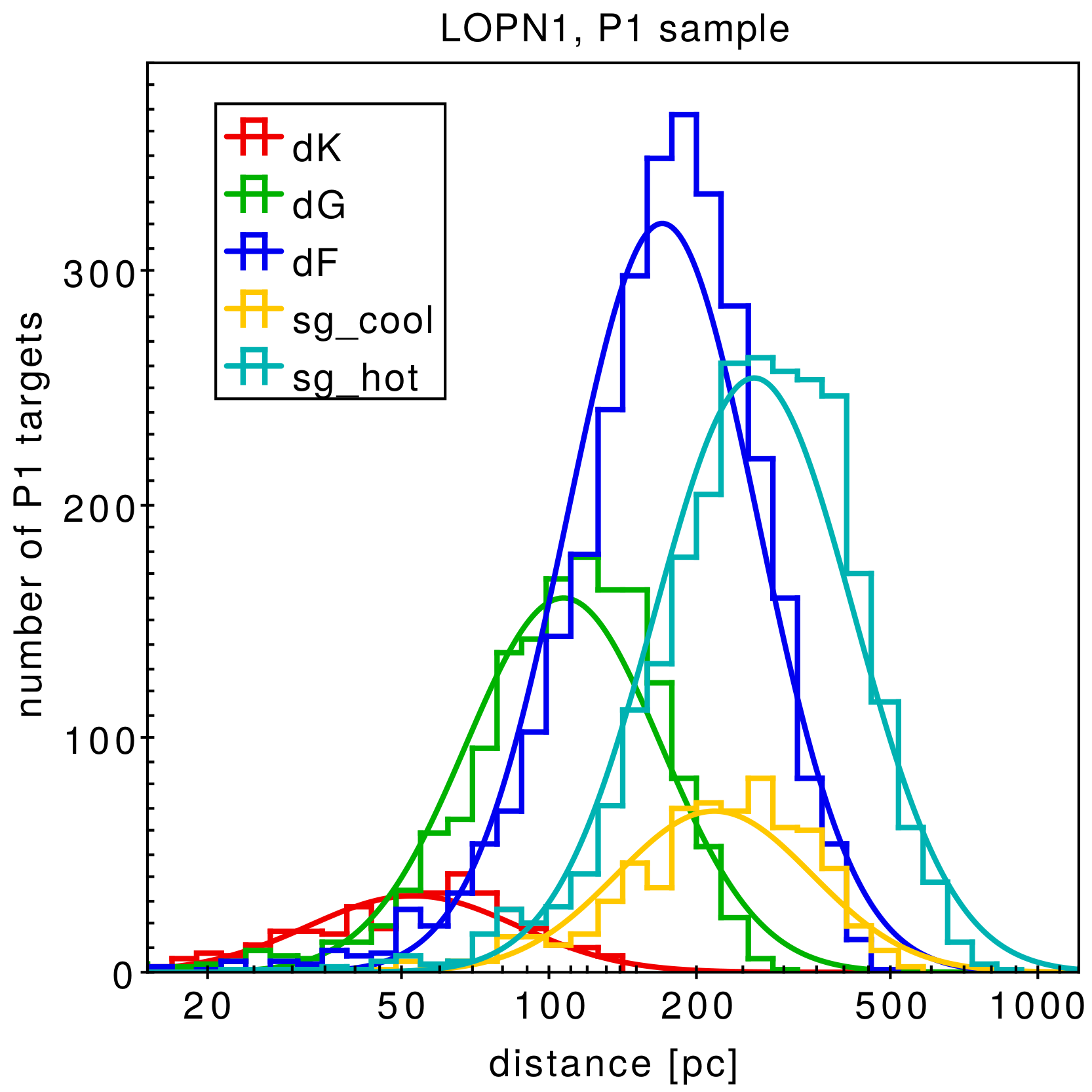} \hspace{1cm}
    \includegraphics[width=0.8\columnwidth]{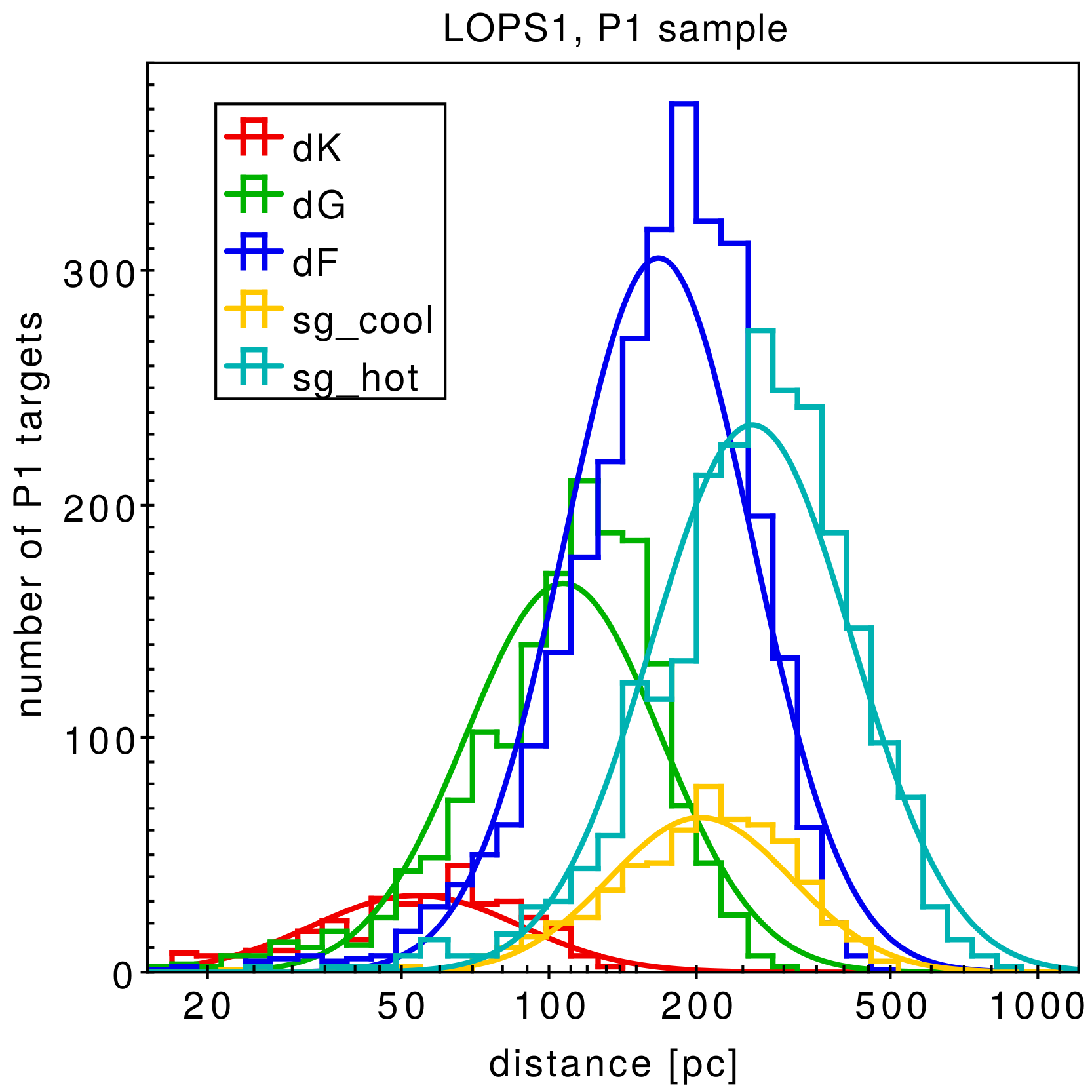}
    \caption{{Astrophysical parameters of the P1 targets} within the LOPN1 (left panels) and LOPS1 (right panels) provisional fields. \emph{Top panels:} Distribution of P1 targets on the $(T_\textrm{eff}, R_\star)$ plane; stars are classified into five distinct groups representing main-sequence K-type dwarfs (\texttt{dK}), G dwarfs (\texttt{dG}), F dwarfs (\texttt{dF}), subgiants at $T_\textrm{eff}<5500$~K (\texttt{sg\_cool}), and $T_\textrm{eff}>5500$~K (\texttt{sg\_hot}). The labels show their relative occurrence. The gray line in both panels marks an arbitrary boundary between main-sequence and evolved stars (see text). \emph{Bottom panels:} Histograms of the distribution in distance of the same subgroups. The best-fit lognormal distribution is also over-plotted with a continuous line.}
    \label{fig:p1content2}\sffamily
\end{figure*}

\subsection{Stellar parameters}\label{content:parameters}

Since for each provisional field we have a catalog containing all the stars suited for inclusion in the P1-P2-P4-P5 samples, it is natural to ask ourselves how the stellar parameters of our targets are distributed, and what selection effects are at work. Again, we mainly focus on the P1 sample as it is recognized as the backbone for the PLATO mission. All the parameters discussed here are extracted from the asPIC and are therefore homogeneously derived on the whole sky based on Gaia DR2 astrometry and photometry and estimated as fully described in \citet{Montalto2021}.
We emphasize that while those model-dependent astrophysical parameters are prone to systematic uncertainties, P1 targets will be thoroughly characterized by PLATO through asteroseismic analysis, yielding stellar radii and ages with an unprecedented degree of accuracy (see Table~1 from \citealt{Aerts2019}; \citealt{Lebreton2014}). Nevertheless, no reasonable systematic offset of the \citet{Montalto2021} parameters (see their Table~4) can significantly change the results of the following analysis.

As a crucial starting point for what concerns the detection and confirmation of planetary transits, we can look at the distribution of our P1 targets as a function of their effective temperature and stellar radius (Fig.~\ref{fig:p1content1}, top panels).
The most obvious feature seen in these number density plots is the ridge corresponding to the main sequence. A long tail of stars extending up to 4-5~$R_\odot$ is made of a well-defined branch of ``typical'' field K- and late-G-type subgiants (at $T_\textrm{eff}\simeq 5000$~K) and, more evenly spread at higher temperatures, of early-G and F subgiant stars. As expected for a bright magnitude-limited selection, our sample is dominated by stars, on average, larger than the Sun and hence with a larger luminosity, with a median of 1.62 and 1.67~$R_\odot$ for the LOPS1 and LOPN1 respectively; this is also a hint that the small numerical gain of P1 targets in the LOPN1 versus~LOPS1 is actually accompanied by a relative increase in the F and subgiant fraction. As for the Galactic latitude distribution of the P1 targets (Fig.~\ref{fig:p1content1}; middle panels), their histograms confirm our expectations that they mostly lie in the $15^\circ\lesssim |b|\lesssim 35^\circ$ range, where the 18- and 24-telescope coverage and the corresponding improvement in photometric precision allows more targets to get through the noise requirements for the P1 sample. Due to the relation between $(T_\textrm{eff}, R_\star)$ and absolute magnitude, the average distance of P1 targets is a strong function of their radius (Fig.~\ref{fig:p1content1}; bottom panels). In particular, while the median distance of 1~$R_\odot$ stars is about 100~pc, the high-luminosity tail extends up to 500~pc and beyond. 

\begin{figure*}[t!]
    \centering\vspace{1cm}
    \includegraphics[width=0.9\columnwidth]{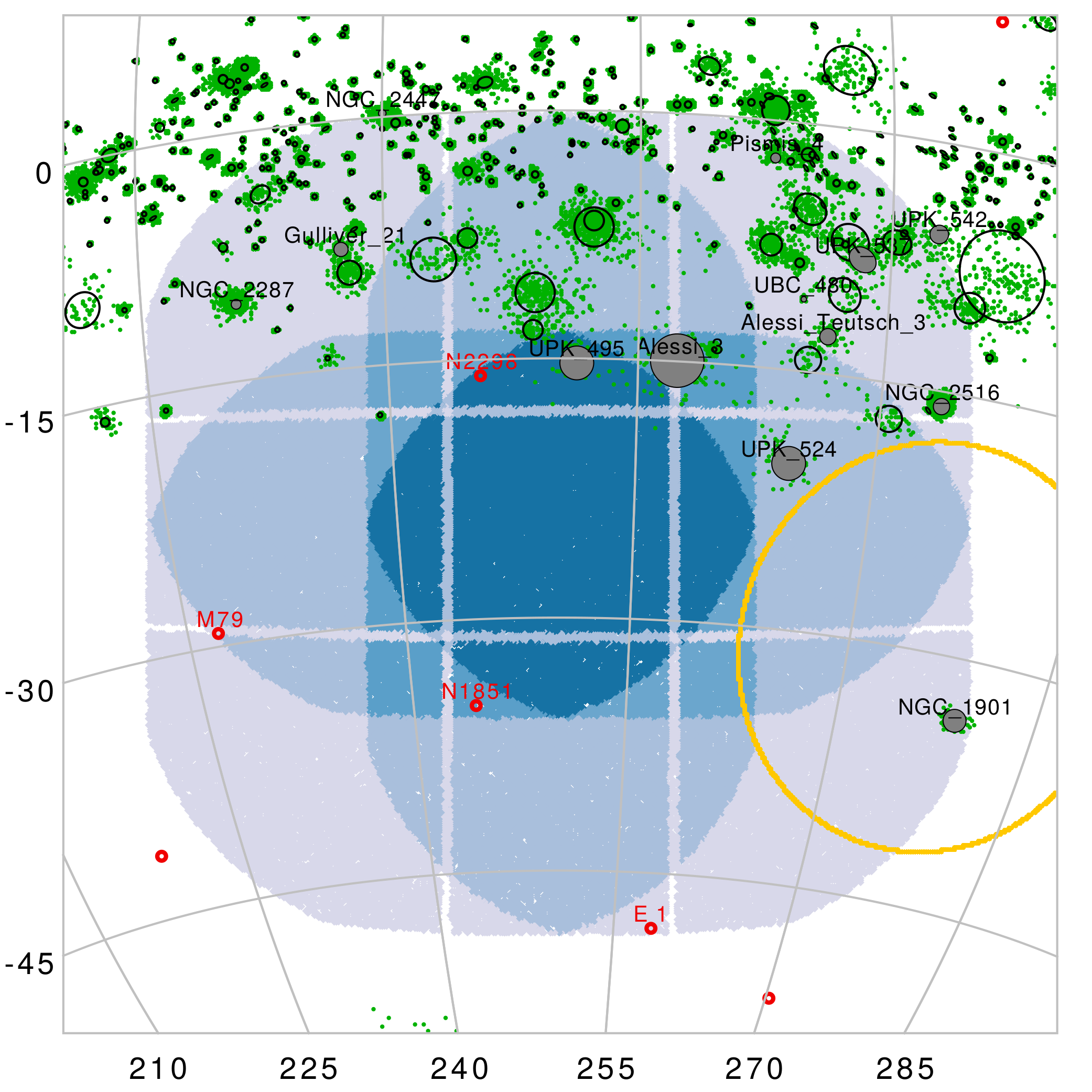}\hspace{6mm}
    \includegraphics[width=0.9\columnwidth]{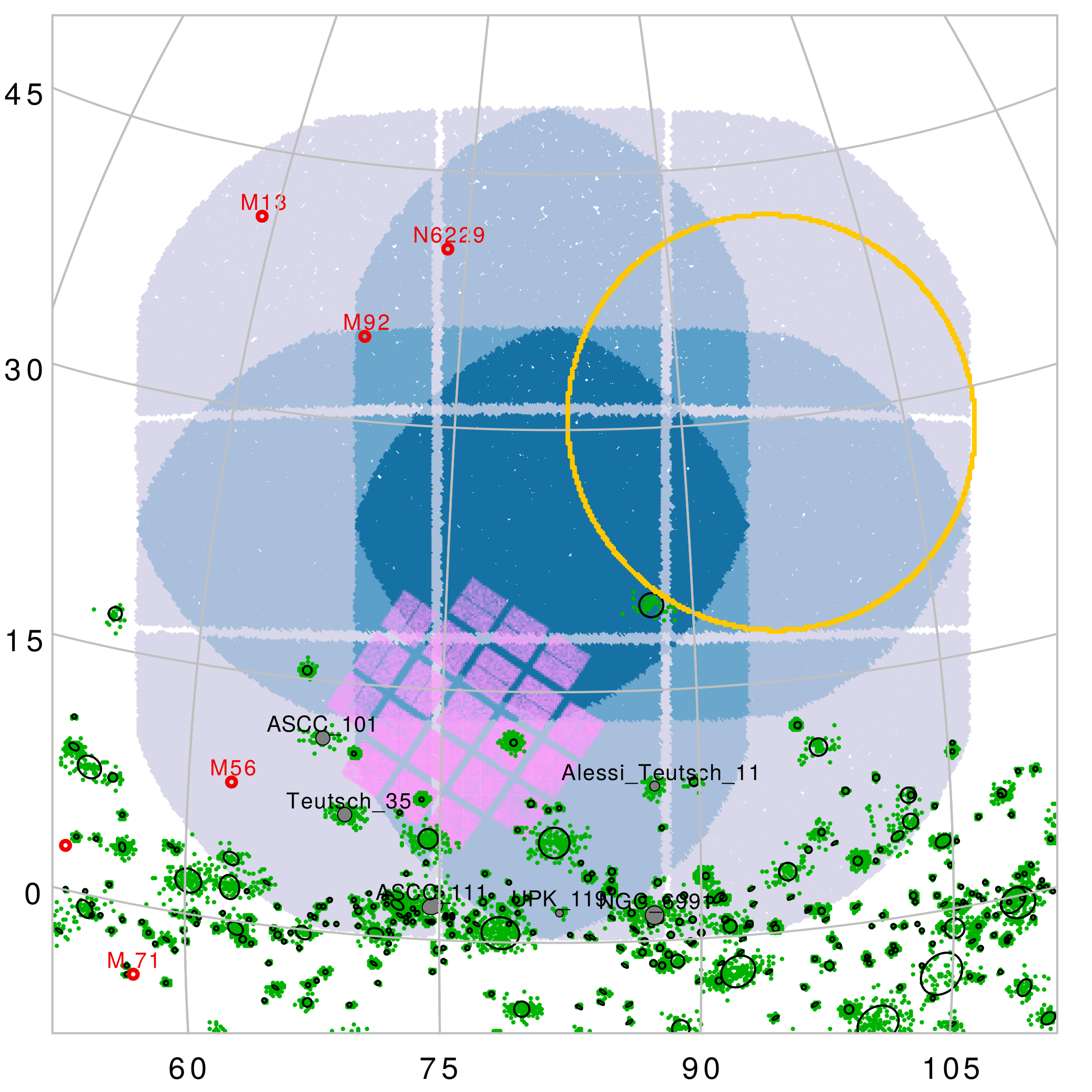} 
    \caption{{Star clusters within the provisional PLATO Fields} LOPS1 (left panel) and LOPN1 (right panel) in Galactic coordinates. The PFs are color-coded in blue shades according to the number of co-pointing ``normal'' cameras, from six (light blue) to 24 (dark blue). The Kepler Field is plotted as a pink translucent area. Open cluster members identified by \citet{Cantat-Gaudin2020} are plotted as green points, and the corresponding cluster center and radius as black circles; only for OCs older than 100~Myr and for which the apparent visual magnitude of a solar twin is $m_{V,\odot} < 15$ ($d\lesssim 1000$~pc) are the circles gray-filled with the name of the OC labeled. Globular clusters from \citet{Harris1996} are also plotted with red empty circles.}
    \label{fig:clusters}
\end{figure*}

In order to dive a bit deeper into the behavior of those different components, we split the LOPS1 and LOPN1 samples into five different subgroups representing main-sequence K-type dwarfs (\texttt{dK}), G dwarfs (\texttt{dG}), F dwarfs (\texttt{dF}), as well as evolved stars on the classical subgiant branch (\texttt{sg\_cool}) and at earlier types (\texttt{sg\_hot}). The $T_\textrm{eff}$ boundaries between \texttt{dK}, \texttt{dG}, and \texttt{dF} are taken from the tables by \citet{Pecaut2013}, the upper limit of the main sequence is parametrized as $R_\star = 1+2\cdot10^{-7}(T_\textrm{eff}-4000)^2$, and the arbitrary boundary between \texttt{sg\_hot} and \texttt{sg\_cool} is set at  5500~K. The position of these five groups on the $(T_\textrm{eff}, R_\star)$ plane and their distance distribution are shown in Fig.~\ref{fig:p1content2} with a consistent color coding. Indeed, about 60\% of the whole sample is made of main sequence stars, with approximately two-thirds of them being F dwarfs (Fig.~\ref{fig:p1content2}, top panels). The overall \texttt{dF}, \texttt{dG}, and \texttt{dK} fraction is about 38\%, 20\%, and 4\%, respectively. The \texttt{dF} and \texttt{sg\_hot} are slightly over-represented in the LOPN1 with respect to LOPS1 at the expense of the \texttt{dG} share. In absolute terms, however, the \texttt{dG} counts are 3.8\% larger in the LOPN1 (1632 versus~1570), which implies that the, on average, lower interstellar extinction at $d>200$~pc in the northern region manages to get some more faint solar analogs into the sample. As for the distance histograms of the five subgroups (Fig.~\ref{fig:p1content2}, bottom panels), they again reflect the luminosity effect, with the intrinsically faint \texttt{dK} component dominating at $d<50$~pc and the \texttt{dF} + \texttt{sg\_cool} + \texttt{sg\_hot} one being the vast majority of targets at $d>200$~pc.

\begin{table*}[t!]\centering
\caption{Properties of open clusters overlapping the provisional PLATO Fields LOPS1 and LOPN1.}
\begin{tabular}{lrrrrrrrr}
\hline
name & $\alpha$ [2000.0] & $\delta$ [2000.0] & $r_{50}$~[deg] & $N_{0.7}$ & $A_V$ & $m_{V,\odot}$ & $d$~[pc] & age~[Myr] \\
\hline\hline
  Alessi 3 & 109.275 & -46.142 & 1.541 & 164 & 0.14 & 12.34 & 297 & 631\\  NGC 2516 & 119.527 & -60.8 & 0.496 & 652 & 0.11 & 13.08 & 423 & 240\\
  ASCC 101 & 288.399 & 36.369 & 0.372 & 69 & 0.19 & 13.10 & 412 & 490\\
  NGC 1901 & 79.561 & -68.294 & 0.652 & 69 & 0.21 & 13.19 & 427 & 891\\
  UBC 480 & 119.954 & -50.636 & 0.094 & 13 & 0.08 & 13.32 & 481 & 209\\
  Teutsch 35 & 294.091 & 35.742 & 0.414 & 143 & 0.00 & 13.37 & 509 & 102\\
  NGC 6991 & 313.621 & 47.4 & 0.555 & 243 & 0.20 & 13.84 & 577 & 1549\\
  UPK 524 & 105.304 & -54.432 & 0.957 & 58 & 0.30 & 13.91 & 569 & 107\\
  UPK 495 & 105.863 & -40.886 & 0.939 & 39 & 0.05 & 13.98 & 660 & 437\\
  NGC 2287 & 101.499 & -20.716 & 0.332 & 625 & 0.02 & 14.04 & 688 & 170\\  Gulliver 21 & 106.961 & -25.462 & 0.364 & 121 & 0.13 & 14.09 & 670 & 275\\
  Alessi Teutsch 11 & 304.127 & 52.051 & 0.257 & 110 & 0.37 & 14.21 & 634 & 145\\
  Pismis 4 & 128.79 & -44.407 & 0.291 & 98 & 0.18 & 14.22 & 695 & 120\\
  UPK 537 & 126.57 & -52.113 & 0.755 & 51 & 0.54 & 14.36 & 628 & 234\\
  Alessi Teutsch 3 & 118.228 & -53.022 & 0.491 & 87 & 0.47 & 14.50 & 693 & 105\\
  UPK 542 & 134.258 & -54.303 & 0.544 & 33 & 0.49 & 14.86 & 808 & 631\\
  ASCC 111 & 302.891 & 37.515 & 0.537 & 136 & 0.39 & 14.87 & 851 & 275\\
  NGC 2447 & 116.141 & -23.853 & 0.202 & 731 & 0.05 & 14.92 & 1018 & 575\\
  UPK 119 & 308.572 & 43.326 & 0.197 & 24 & 0.71 & 14.96 & 765 & 1549\\
  \hline\end{tabular}\label{table:clusters}
  \tablefoot{The columns show the cluster name, the equatorial coordinates $\alpha$ and $\delta$ in decimal degrees, the angular radius $r_{50}$ enclosing 50\% of the cluster members, the number $N_{0.7}$ of cluster members having a membership probability larger than 0.7, the extinction coefficient $A_V$, the $V$-band apparent magnitude $m_{V,\odot}$ of a cluster star having the Sun's absolute magnitude $M(V)=4.83$, the distance $d$ in parsecs, and the best-fit age in Myr. The table is sorted by increasing $m_{V,\odot}$. All the parameters except for $m_{V,\odot}$ are extracted from \citet{Cantat-Gaudin2020}.}
\end{table*}

\subsection{Star clusters}\label{content:clusters}

Star clusters are known as excellent astrophysical laboratories because their ensemble properties (including age, distance, and chemical composition) can be measured or constrained more precisely than is possible with single field stars. This holds not only for stellar astrophysics, but also in exoplanetology, where clusters provide us with a homogeneous environment where the physical and orbital parameters of the discovered planets can be put into an accurate evolutionary context \citep{Dawson2018}. The large pixel scale of PLATO makes the extraction of high-precision light curves in crowded environments very challenging; nevertheless, sophisticated  difference image analysis (DIA) and PSF-fitting techniques have been developed to overcome this issue \citep{Bouma2019,Montalto2020,Nardiello2020}, so it is worth asking which clusters will be imaged during the LOP phase. This is also requested by a SRD science requirement, which discusses the need to use PLATO to investigate planet formation in different environments, including clusters.

We start from the work by \citet{Cantat-Gaudin2020}, who list model-dependent astrophysical parameters and membership probabilities for more than $200\,000$ stars hosted by 1867 confirmed open clusters (OCs) and young associations. About 300 OCs overlap at least partly with the LOPS1 and LOPN1 footprint (Fig.~\ref{fig:clusters}). With only a handful of exceptions, they are located quite close to the Galactic disk at $|b|<15^\circ$ and simultaneously monitored by six or 12 PLATO cameras. Discussing the single clusters is beyond the scope of this paper, but it is indeed interesting to focus on OCs being particularly favorable to be targeted by PLATO, that is, close enough for solar-type stars to be below a reasonable magnitude limit, and not so young for stellar activity to be a severe limiting factor. From cluster distance $d$ and extinction coefficient $A_V$, we can easily compute the apparent visual magnitude $m_{V,\odot}$ of a hypothetical solar twin, assuming an absolute magnitude $M_{V,\odot}\simeq 4.83$. Then, by setting $m_{V,\odot} < 15$ ($V\sim 15$ being an approximate upper limit for both the PLATO photon-noise-dominated regime and for an effective RV follow-up of giant planet candidates; roughly corresponding to $d\lesssim 1000$~pc) and age~$>100$~Myr, we get 19 OCs, 13 in the LOPS1 and six in the LOPN1, listed in Table~\ref{table:clusters} and over-plotted with labeled gray circles in Fig.~\ref{fig:clusters}. We stress that those cluster ages, in particular, are prone to large systematic errors and are considered here just to set a rough boundary between very young clusters and older ones. Notably most of the selected OCs, although their nature of bound stellar systems is confirmed, are scarcely characterized; with the notable exceptions of NGC~2516 \citep{Fritzewski2020,Bouma2021}, NGC~1901 \citep{Carraro2007}, NGC~2287 \citep{Harris1993,Sun2019}, and NGC~2447 \citep{Reddy2015,DaSilveira2018}, all of them have never been investigated through high-resolution spectroscopy or by targeted photometric surveys.

An even harder challenge for PLATO is the extraction of light curves of stars belonging to globular clusters (GCs) due to their larger distance and extreme crowding. While the PSF-based approach has been demonstrated to be effective in this regime even with smaller telescopes and larger pixel scales, such as with TESS \citep{Nardiello2019}, clearly the limiting magnitude prevents us from reaching the lower main sequence. Rather than for exoplanet discovering, such data could be very useful to carry out rotational and asteroseismological studies of their giant stellar population. Both LOPS1 and LOPN1 are angularly quite far from the Galactic bulge, where the vast majority of GCs are. Still, by cross-matching the \citet{Harris1996} catalog with the field footprints, we get eight GCs imaged during the LOP phase (labeled red points in Fig.~\ref{fig:clusters}), four in the LOPS1 (AM~1=E~1, M~79, NGC~1851, and NGC~2298), and four in the LOPN1 (M~13, M~56, M~92, and NGC~6229). Three of them will be monitored by 18 or 24 telescopes: M~92, NGC~1851, and NGC~2298. 
As an additional note, we mention that about half of the Large Magellanic Cloud (LMC; yellow circle in Fig.~\ref{fig:LOPS1planets}) will be observed in LOPS1 by six PLATO cameras, complementing and extending the ongoing monitoring by TESS \citep{Sharma2018,Mackereth2021}.

\subsection{Eclipsing binaries}\label{content:ebs}

Detached eclipsing binaries (DEBs) have long been studied since the
masses and radii of both stellar components can be determined to good
accuracy, which can then be used to test stellar models
\citep{Popper1967,Popper1980,Torres2010,Serenelli2021}. However, as the binary orbit
has to be sufficiently wide so that the stars evolve as if they were
single stars, this restricts binaries that have an orbital period of
a day or more. The latest version (September 6, 2021) of \texttt{DEBCAT}
\citep{Southworth2015}, which is a continuation of the catalog of
\citet{Andersen1991}, lists 271 DEBs.

Since the ephemerides of DEBs are typically well known, they can be
used as verification sources for the systematic accuracy of the  PLATO time stamps. In addition, as \citet{Maxted2018} argues, they can also be used to validate the PLATO asteroseismic age scale. Doing a cross-match of {\tt DEBCAT} with the footprint of fields LOPS1 and LOPN1 yields 21 and 45 DEBs, respectively. Among these, eight and 12 (respectively) meet the magnitude, spectral type, and noise requirements to be considered for inclusion in the P1-P2-P4-P5 stellar samples.

\subsection{Synergies}\label{content:synergies}

There is a considerable overlap between LOPS1, LOPN1, and other past and future space-based missions. Notably, the Kepler Field is fully within the footprint of LOPN1 (Fig.~\ref{fig:clusters}, right panel). About 35\% of its targets will be imaged by either 24 or 18 PLATO cameras (in particular at $b\gtrsim 13^\circ$), 58\% by 12, and 7\% by six (that is, by a single group of telescopes). Among the $200\,036$ targets listed in the Kepler Stellar Properties Catalog DR25 \citep{Brown2011}, $12\,418$ stars are formally included in the PLATO P1-P2-P5 samples (FGK dwarfs subgiants), with $\sim{}700$ M dwarfs meeting the requirements of the P4 sample. This is mostly due to the magnitude constraint of PLATO which is much brighter than the median magnitude of the Kepler sample at $\textrm{Kepmag}\simeq 14.6$. Nevertheless, as mentioned in Section~\ref{content:exoplanets}, all the stars with confirmed planets for which meaningful photometry can be extracted will be included in the PIC irrespective of their formal compliance with the requirements. 

TESS finished its two-year nominal mission in July 2020, during which its scanning law defined two CVZs at $|\beta|>78^\circ$ each covered by 13 sectors, or one year of mostly continuous photometry. During the first extended mission, Year 3 has followed a similar pattern as Year 1 (thus re-observing the southern ecliptic hemisphere) while, since August 2021, Year-4 TESS observations started mapping part of the ecliptic region for the first time.  While the actual observing strategy of the next extended missions has not been officially chosen yet, it is clear that the ecliptic caps will continue to retain a crucial importance in terms of phase coverage given the mission design. Even more important, the ecliptic polar regions are in strong synergy with other follow-up and characterization missions that will be flying during or after the PLATO nominal mission. In addition to JWST, another satellite will be extremely effective in characterizing the atmospheres of planets discovered by PLATO, especially those hosted by bright and nearby P1 stars: ARIEL \citep{Tinetti2018}, whose CVZ is sligthly larger than the TESS one ($|\beta|\gtrsim 70^\circ$; \citealt{Tinetti2021}). The north TESS CVZ is fully within LOPN1 (Fig.~\ref{fig:clusters}, right panel), and about 75\% of it is covered by 12 or more telescopes; among the 361,196 unique candidate target list (CTL) targets observed by TESS from sector 1 to 42, 16,511 meet the requirements to be considered for inclusion in the PLATO sample P1 or P5 (4\,882 for P1 alone); in other words, $\sim 60\%$ of the LOPN1 P1 targets already have TESS short-cadence data. As for LOPS1 (Fig.~\ref{fig:clusters}, left panel), more than half of the $\beta<-78^\circ$ CVZ is secured, mostly in the six- and 12-telescope region. Including a larger fraction of it would imply moving LOPS1 too close to the crowded Galactic plane, dramatically decreasing our prioritization metric (Fig.~\ref{hpix4}, right plot). Nevertheless, among the same TESS CTL targets mentioned above, 16,881  meet the requirements to be considered for inclusion in the PLATO sample P1 or P5 (5\,195 for P1 alone, or $\sim 67\%$ of the whole LOPS1 P1 sample). The mean number of available TESS sectors is currently 14.1 for LOPN1 P1 targets and 13.8 for LOPS1 P1, with the shorter time spent by TESS on the north Ecliptic cap almost exactly balancing the smaller overlap between LOPS1 and the TESS CVZ.

As a closing remark, we note that LOPS1 (but not LOPN1) crosses a large sector of the $40^\circ\lesssim |\beta|\lesssim 50^\circ$ band where the scanning law of Gaia \citep{Gaia2016} will result in the largest number of scanning transits and hence much better astrometric solutions, more reliable astrophysical parameters, and also more densely sampled time-resolved measurements. For the same reason, this is also the sweet spot where Gaia is supposed to discover a large number of planetary systems through the astrometric technique \citep{Sozzetti2014}. Although only a tiny fraction of them will be on transiting configurations, there is a chance that PLATO will be able to catch some of them. In any case, PLATO will contribute to better characterize the architecture of some of these systems providing transits of internal planets (when orbit inclination will be favorable).

\section{Conclusions}

The PLATO launch date is slowly approaching, being currently planned by the end of 2026.  Two years before the launch, a formal proposal for the first field to be observed (likely a LOP field) has to be delivered, and the final PIC for the first field shall be frozen nine months before launch to allow, among other things, the kick-off of the guest observer (GO) call for proposals. 

In this work, we have outlined the complex problem concerning the PLATO field selection in general, discussed the many different (and sometimes competing) scientific criteria involved, and developed a prioritization metric to help the process, which unavoidably is a multi-staged one: The current provisional LOP fields we presented, named LOPS1 and LOPN1, and of which we described content and synergies, will be fine-tuned in the next years. It is likely that the final choice will not move from the current proposal by more than a few degrees (cf.~the $3.7^\circ$ average grid spacing in Fig.~\ref{hpix4}). For such a huge FOV of $>2000$~$\textrm{deg}^2$, that would translate into just a few percent of re-allocated targets. 
There is still time for the final choices of the fields PLATO will observe, and one of the purposes of this paper is to stimulate the  scientific community to provide inputs to help converging into the final choice of LOPN and LOPS.

\begin{acknowledgements}

The authors are grateful to the referee, Keivan Stassun, for reading the manuscript and providing constructive comments and suggestions.
This work presents results from the European Space Agency (ESA) space mission PLATO. The PLATO payload, the PLATO Ground Segment and PLATO data processing are joint developments of ESA and the PLATO Mission Consortium (PMC). Funding for the PMC is provided at national levels, in particular by countries participating in the PLATO Multilateral Agreement (Austria, Belgium, Czech Republic, Denmark, France, Germany, Italy, Netherlands, Portugal, Spain, Sweden, Switzerland, Norway, and United Kingdom) and institutions from Brazil. Members of the PLATO Consortium can be found at \url{https://platomission.com/}. The ESA PLATO mission website is \url{https://www.cosmos.esa.int/plato}. We thank the teams working for PLATO for all their work.
MM, GP, VN, VG, LP, SD, SO, SB, RC, LM, IP acknowledge support from PLATO ASI-INAF agreements n.2015-019-R0-2015 and n. 2015-019-R.1-2018. 
We would like to acknowledge the financial support of INAF (Istituto Nazionale di Astrofisica), Osservatorio Astronomico di Roma, ASI (Agenzia Spaziale Italiana) under contract to INAF: ASI 2014-049-R.0 dedicated to SSDC. 
CA acknowledges funding received from the KU\,Leuven Research Council (grant C16/18/005: PARADISE) and from the BELgian federal Science Policy
Office (BELSPO) through PRODEX grants for Gaia and PLATO.
This work has made use of data from the European Space Agency (ESA) mission
{\it Gaia} (\url{https://www.cosmos.esa.int/gaia}), processed by the {\it Gaia}
Data Processing and Analysis Consortium (DPAC,
\url{https://www.cosmos.esa.int/web/gaia/dpac/consortium}). Funding for the DPAC
has been provided by national institutions, in particular the institutions
participating in the {\it } Multilateral Agreement.
This research has made use of the SIMBAD database (operated at CDS, Strasbourg, France), TOPCAT and STILTS \citep{Taylor2005,Taylor2006}.

\end{acknowledgements}

\bibliographystyle{aa}
\bibliography{biblio}

\end{document}